\documentclass[fleqn,usenatbib]{mnras}

\usepackage{newtxtext,newtxmath}
\usepackage[T1]{fontenc}
\usepackage{ae,aecompl}
\usepackage{multirow}
\usepackage{tabularx}
\usepackage{supertabular}
\usepackage{longtable}

\usepackage[]{url}
\usepackage{graphicx}	% Including figure files
\usepackage{amsmath}	% Advanced maths commands
\usepackage{amssymb}	% Extra maths symbols

\usepackage{graphicx}	% Including figure files
\usepackage{amsmath}	% Advanced maths commands
\usepackage{amssymb}	% Extra maths symbols
\usepackage{multirow}
\usepackage[left]{lineno}
\usepackage{booktabs}
\usepackage{gensymb}
\usepackage{xcolor,color}

\usepackage{csvsimple}
\definecolor{DT_green}{RGB}{0, 204, 102}

\title[GW follow-up using galaxies $M_{*}$]{Optimising gravitational waves follow-up using galaxies stellar mass}

\author[]{
\newauthor
J. -G. Ducoin$^{1}$,
D. Corre$^{1}$,
N. Leroy$^{1}$,
E. Le Floch$^{2}$
\\
$^{1}$LAL, Univ Paris-Sud, CNRS/IN2P3, Orsay, France\\
$^{2}$IRFU, CEA, Univ Paris-Saclay, Gif-sur-Yvette, France}

\date{Accepted 2020 January 10. Received 2020 January 9; in original form 2019 November 11}

\hypersetup{draft}

\begin{document}
\label{firstpage}
\pagerange{\pageref{firstpage}--\pageref{lastpage}}
\maketitle

% Abstract of the paper
\begin{abstract}

We present a new strategy to optimise the electromagnetic follow-up of gravitational wave triggers. This method is based on the widely used galaxy targeting approach where we add the stellar mass of galaxies in order to prioritise the more massive galaxies. We crossmatched the GLADE galaxy catalog with the AllWISE catalog up to 400Mpc with an efficiency of $\sim$93\%, and derived stellar masses using a stellar-to-mass ratio using the WISE1 band luminosity. We developed a new grade to rank galaxies combining their 3D localisation probability associated to the gravitational wave event with the new stellar mass information. The efficiency of this new approach is illustrated with the GW170817 event, which shows that its host galaxy, NGC4993, is ranked at the first place using this new method. The catalog, named Mangrove, is publicly available and the ranking of galaxies is automatically provided through a dedicated web site for each gravitational wave event.
\end{abstract}

% Select between one and six entries from the list of approved keywords.
% Don't make up new ones.

%%%%%%%%%%%%%%%%%%%%%%%%%%%%%%%%%%%%%%%%%%%%%%%%%%

%%%%%%%%%%%%%%%%% BODY OF PAPER %%%%%%%%%%%%%%%%%%

\section{Introduction}
%\subsection{Gravitational waves follow-up}

Gravitational waves from binary neutron star (BNS) coalescence, in association to short gamma-ray burst, opened a new era of multi-messenger astronomy. The event of the 17th august 2017 was a real breakthrough for the multi-messenger astronomy. For the first time, an electromagnetic counterpart of a gravitational wave was observed \citep{LSC_BNS_2017PhRvL}. The association with a gamma ray burst (GRB) detected by Fermi-GBM a few seconds after the coalescence provides the first evidence of a link between BNS merger and short gamma ray burst.
The huge effort of ground telescopes follow-up in association to the relative small volume of this event localisation in the sky, allowed the identification of the GW electromagnetic counterpart. The multi-wavelength observations improved our understanding of the physics of strong-gravity and put some constraints on astrophysical models related to matter during the merger and post-merger phase. A first constraint of the speed of gravitational waves and violation of Lorentz invariance has been determined from this event \citep{LSC_GW_GRB_2017ApJ}. Both kilonova and afterglow observations provide information about the neutron star equation of state, energy of the ejecta, merger remnant, ambient medium and so on \citep{Metzgerkilo,2018PhRvL.121p1101A,2019arXiv190906393H,2019ApJ...876..139G}. Such event also provide a new independent derivation of the Hubble Constant \citep{gwtohubble1,gwtohubble2,gwtohubble3}.\\
The third LIGO-Virgo run (O3) started the first of April 2019 and multi-messenger astronomy related to gravitational waves with it. With improved sensitivity of the LIGO-Virgo detectors, the year-long third observing run (O3) has already brought his share of merging binaries and promises many more of them. Therefore an intensive multi-wavelength follow-up of those events with ground and space instruments is performed all around the world. But the identification of the electromagnetic counterpart of such event is very challenging knowing the wide sky localization area provided by LIGO-Virgo (which can span more than 1000 deg$^{2}$ ) and require complex observation strategies.\\
Many efforts were done recently to optimise the observations for these large sky areas \citep{2016A&A...592A..82G, Coughlin19_opt}. For large field of view (FoV) telescope the standard approach consists in observing the localisation error box provided by LIGO-Virgo \citep{bayestar, LALInference} using an optimised tiling of the sky \citep{gwemopt}. In such standard strategy, the scheduling of the tile observation is provided by the 2D probability distribution from LIGO-Virgo skymaps.\\
As suggested by \citep{2016ApJ...820..136G} more recent works tried to include galaxies population to the strategy \citep{LosC, Antolini2017, Rana2019}. Such developments allow to use the 3D probability distribution from the LIGO-Virgo skymap (and not only the 2D probability) to produce a "galaxy weighting" rank for the tiles. It also allows to provide a scheduling of observation for narrow FoV telescopes. Indeed, with such information we can provide a list of interesting galaxies (i.e. ranked by their 3D position probability inside the LIGO-Virgo skymap) to be observed by these small FoV telescopes.
To allow the use of information related to galaxies, one must rely on a galaxy catalog that is sufficiently complete compared to the interferometers sensitivity range. The current binary neutron star range is at 130 Mpc for LIGO Livingston, 110 Mpc for LIGO Hanford, and 50 Mpc for Virgo \citep{BNSrange}. The CLU catalog \citep{CLUcatalog} for the north hemisphere contains the WISE1 luminosity information as well as spectroscopic measurement of local galaxies, however it is non publicly available. Therefore, for the purpose of our work, we rely on the publicly available GLADE galaxy catalog \citep{GLADEcatalog} which is all-sky and complete up to 100Mpc, and nearly complete up to 150Mpc.\\

Given the large size of error boxes, the number of galaxies compatible with an event can be very large (>few thousands). In such cases, the classification using the 3D probability only is limited because it will produce similar values for a large number of galaxies. Adding galaxy properties to the ranking is a way to reduce the sample size of interesting galaxies. Among the various galaxies properties that could influence the rate of BNS merger, such as star formation rate (SFR), stellar mass and metallicity, several works pointed out a significant dependence to the stellar mass \citep{2019MNRAS.487.1675A, 2019MNRAS.tmp.2085T, 2018MNRAS.481.5324M}. Furthermore, short GRB host galaxies are known to be associated to BNS merger since GW170817, and are found in massive galaxies. The short GRB host galaxies are more massive than the long GRBs host galaxies, pointing to the importance of the stellar mass in determining the rate of short GRBs \citep{Leibler2010, Fong2013, Berger2014}. So far, the only one known (by gravitational waves) host galaxy of BNS merger is NGC4993 from GW170817 event. This galaxy present a very high mass and a low star formation rates \citep{NGC4993mass,Blanchard2017,Hjorth2017,Levan2017,Pan2017}. The host galaxy of GRB150101B presented as a analogue of GRB170817A \citep{GRB150101B} is also a vary massive galaxy. In the light of those information we chose to focus on the stellar mass for the selection of gravitational waves host galaxy candidates.\\
\\
In Section \ref{section:gal_target}, we describe the general galaxy targeting approach and the new formulation we propose to include the stellar mass information. In the Section \ref{section:stellarmass} we describe the crossmatch between the GLADE and AllWISE catalogs and the stellar mass estimation as well as an estimation of the completeness of the resulting Mangrove (Mass AssociatioN for GRavitational waves ObserVations Efficiency) catalog in terms of stellar mass. In Section \ref{section:GW170817} we test our method on the GW170817 event. In Section \ref{section:conclusion} we discuss future development of this method for wide field of view telescopes. Throughout this paper, we use the Plank 2015 cosmological parameters \citep{planck15}. 

\section{Galaxy targeting method}
\label{section:gal_target}
In this section, we first describe the standard galaxy approach using the 3D localisation of the GW skymap, and then we propose a new formulation of the grade used to rank galaxies in order to include the stellar mass information.
\subsection{Standard approach}
In case of a gravitational wave event, LIGO-Virgo rapidly releases a probability skymap based on the distance and two dimensional localisation of the event allowing to constrain the region of the sky to search for the GW electromagnetic counterpart \citep{bayestar}. With such skymap we are able to fetch the probability density per unit of volume at a given position \citep{GoingtheDistance}. This is used to infer the probability of a given galaxy to be the host of the merger according to its celestial position $P_{pos}$ with the following relation:
\begin{equation}
\label{eq1}
    P_{pos} = P_{dV} = \frac{P_{pixel}}{Pixel\: area} \: N_{pixel}\: e^{- \frac{1}{2} \left( \frac{D_{galaxy}-\mu_{pixel}}{\sigma_{pixel}} \right)^2}
\end{equation}
\\
Where $P_{pixel}$ is the 2D probability included in the given pixel, $N_{pixel}$ is the normalisation factor for the given pixel, $\mu_{pixel}$ is the mean distance value at the given pixel, $\sigma_{pixel}$ is the standard deviation at the given pixel and $D_{galaxy}$ is the luminosity distance of the galaxy fetched from the galaxy catalog.
The outputs of the LIGO-Virgo localization pipelines are HEALPix (Hierarchical Equal Area isoLatitude Pixelization) all-sky images, the skymap we are dealing with is composed of pixels defined by the HEALPix \citep{HEALPix} format.\\
For the selection of the galaxies, we classified as "compatible" with the skymap, a galaxy which fulfills the two following conditions:
\begin{itemize}
    \item Its 2D position in the sky as to be in the $90\%$ of the 2D skymap probability distribution.
    \item Its distance has to fall within the 3 sigma distance error localization at the given pixel of the galaxy.
\end{itemize}

\begin{figure}
\begin{center}
\includegraphics[width=1.0\columnwidth]{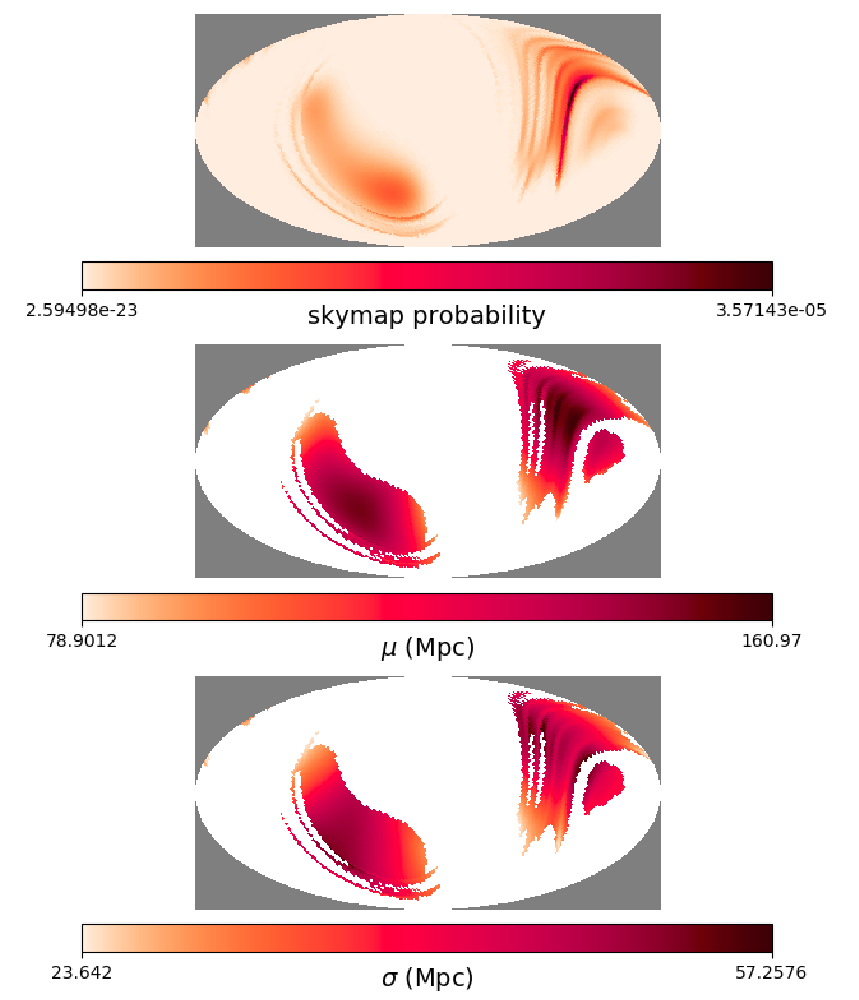}
\caption{The top plot show the probability distribution for the BNS candidate S190425z of O3. The middle and bottom plot are respectively the mean distance and its standard deviation distributions for the pixels inside the the $90\%$ of the 2D skymap probability distribution}
\label{fig:meandist}
\end{center}
\end{figure}

\begin{figure}
\begin{center}
\includegraphics[width=1.0\columnwidth]{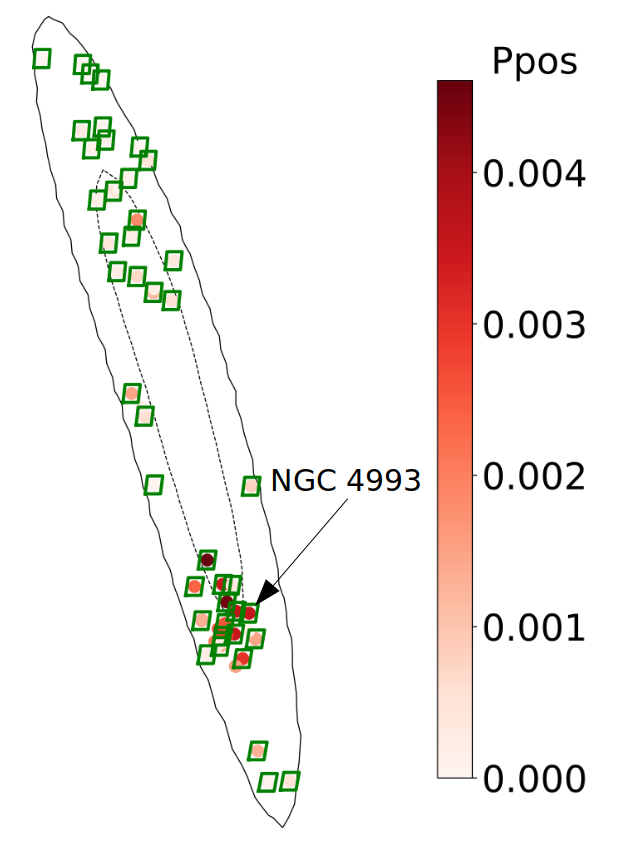}
\caption{Skymap of GW170817. The dashed and the solid line enclose respectively the $50\%$ and the $90\%$ of the skymap. The green squares represent the pointing of a telescope with a field of view of 20'. The red color scale show the grade of the galaxies according to (\ref{eq3}). The black arrow show the position of the GW host galaxy, NGC 4993.}
\label{fig:gw170817galtar}
\end{center}
\end{figure}

With such conditions we ensure that telescopes will not point outside of the $90\%$ skymap probability distribution. The conservative choice of 3$\sigma$ on the distance constraint is motivated by the fact that galaxies with a low distance probability will be always penalised in the ranking process. Regarding the condition on the distance, we use the pixel by pixel information and not the mean distance estimation for the LIGO-Virgo candidate, as in the approach adopted by \citep{LosC}. This is more efficient because the distance estimation can be very inhomogeneous in a given skymap. The figure \ref{fig:meandist} shows the distribution pixel by pixel of the mean distance $\mu$ and the standard deviation $\sigma$ inside the the $90\%$ of the 2D skymap probability distribution for the S190425z candidate during the O3 run. This example of a BNS candidate shows that for a large portion of the skymap the $\mu$ and $\sigma$ at a given pixel are far away from the mean distance evaluation for the whole skymap (155$\pm$ 45 Mpc in this case). Using pixel by pixel information prevents to dismiss compatible galaxies or select incompatible ones.
One of the main advantage of the galaxy targeting is to reduce the amount of observations necessary to cover a given skymap.% Assuming the completeness of the galaxy catalog and knowing that the source is related to a galaxy you can consider a skymap as fully observed if you observed all the galaxies compatible with it.
The figure \ref{fig:gw170817galtar} shows the result of the galaxy targeting for GW170817 with a standard field of view of 20 arcmin. A large part of the skymap is not worth observing because it does not contain any compatible galaxies. A total of 44 pointing are needed to observe all galaxies compatible with the $90\%$ skymap whereas it would have required 144 tiles to cover the entire $90\%$ skymap, which results on significant gain of observational time allocation and revisits.
\\
\\
In such a standard strategy only the 3D position of the galaxy is used, and a more advanced strategy is to include physical properties of the galaxies such as the stellar mass as the host galaxy of BNS counterparts are likely to be found in massive galaxies.

%In the next sections, we will describe how to take into account the stellar mass for the GLADE galaxies using their WISE1 band luminosity, and then how to include it in the galaxy ranking process.

\subsection{Grade reformulation with stellar mass}
\label{subsection:Gradereformulation}
In this section, we describe how to take into account the stellar mass information in the galaxy ranking process, in order to prioritize massive galaxies. We introduce a new term $G_{mass}$ defined as:
\begin{equation}
\label{eq2}
    G_{mass} = \frac{M_{*,galaxy}}{\sum{M_{*,galaxy}}}
\end{equation}
where $M_{*,galaxy}$ is the stellar mass of a given galaxy and the sum is over all of the galaxies compatibles with a given skymap (see \ref{section:gal_target} for the definition). This We first combine this term to the standard grade defined in equation (\ref{eq1}) for each galaxy computing the product as:
\begin{equation}
\label{eq3}
    G_{tot} = P_{pos} \times G_{mass}
\end{equation}
The equation (\ref{eq3}) is proposed to fit with \citep{LosC} expression to allow a direct comparison of the results (see Section \ref{section:GW170817}), but the drawback of this expression with a simple product is that galaxies for which no stellar mass is available are simply not considered. 
In order to keep galaxies without stellar mass information, and still use their 3D localisation probability, $P_{pos}$ from equation (\ref{eq1}), we propose to redefine the grade defined in equation (\ref{eq3}) as:
\begin{equation}
\label{eq4}
    G_{tot} = P_{pos} \thinspace (1 + \alpha \beta G_{mass})
\end{equation}
where $\alpha$ and $\beta$ are positive real parameters. With such definition, $G_{mass}$ is set to 0 when the stellar mass information is not available to fall back on $P_{pos}$. The parameter $\alpha$ is defined such that the two terms in equation (\ref{eq4}) contribute equally to the total grade, $G_{tot}$:
\begin{equation}
\label{eq5}
    \frac{\sum{P_{pos}}}{N} = \frac{\sum{P_{pos}\thinspace \alpha \thinspace G_{mass}}}{N}
\end{equation}
\begin{equation}
\label{eq6}
    \Rightarrow \alpha = \frac{\sum{P_{pos}}}{\sum{P_{pos} \thinspace G_{mass}}}
\end{equation}
where N is the total number of galaxies compatible for a given skymap having a determined stellar mass, and the sum is also over all galaxies compatible for a given skymap having a determined stellar mass. The parameter $\beta$ is used to weight the importance of $G_{mass}$ in the total grade, it is skymap independent. Ideally, $\beta$ should be fitted on a statistically significant sample of gravitational wave host galaxies, but as only one event has been detected so far, we simply chose to put $\beta$ equal to one. \\
\\
Previous works (\cite{LosC,WaW}) chose to include an other factor of the grade which describe the likelihood to detect the counterpart according to limiting magnitude of the observing telescope and the expected magnitude of the source. 
Such factor can be added to the expression (\ref{eq3}) and (\ref{eq4}) if needed. We choose not to develop such strategy because first the limiting magnitude of a telescope can vary a lot between two observations (seeing, horizon...) and secondly only one detection of gravitational wave electromagnetic counterpart has been achieved at the moment, so only one one set of data describing the expected kilonova lightcurve is available and assuming a standard lightcurve on a single object could be risky.\\
\\
The reformulation of the grade presented here is independent of the method used to determine the stellar mass of galaxies. In section \ref{section:stellarmass} we present the method we have chosen to follow to obtain the stellar mass of galaxies in a homogeneous way for a catalog over the whole sky. The resulting catalog is fully publicly available as described in section \ref{section:conclusion}

\section{Adding stellar mass to GLADE catalog}
\label{section:stellarmass}
In this section we aim at adding the stellar mass for the GLADE galaxy catalog. In the first part, we describe how to derive the stellar mass from the WISE1 band luminosity. In the second part we study the completeness of the Mangrove catalog resulting from the crossmatch between GLADE and AllWISE catalogs.

\subsection{WISE1 band as a prob of stellar mass}
\label{subsection_match}

Homogeneity in the stellar mass estimation is crucial for our purpose as it is used to rank the galaxies. Unfortunately, the GLADE galaxy catalog we rely on provides the B, J, H and K band magnitudes for some of the galaxies but not the stellar mass. We did not chose to compile the stellar masses from various works for homogeneity reasons as it would bring systematics in the ranking process due to possibly quite different methods in the stellar mass estimation. Previous work \citep{LosC} used the B band magnitude as an indicator of the stellar mass. However, the B band is sensitive to the star formation history and can be strongly affected by dust extinction \citep{dustext2, FM07}. The near-infrared luminosity emitted by the old stellar population is fairly insensitive to dust extinction, and is thus considered as a reliable indicator of the total stellar mass of a galaxy \citep{Bruzual2003,Maraston2005}. In our work, we restrict the distance to 400Mpc as it is reasonably above the limiting sensibility distance of LIGO-Virgo for binary neutron star merger with O3 sensibility \citep{BNSrangecalc}. Up to 400 Mpc, there are only $\sim$67\% of the galaxies with a K band magnitude in the version 2.3 of the GLADE catalog. As showed by \citep{MLWISE11, MLWISE12}, the WISE1 (3.4 $\mu$m) luminosity is a reliable indicator of the stellar mass, moreover the AllWISE catalog \citep{AllWISE} has the advantage to be an all-sky catalog. Therefore we made the choice to perform a crossmatch between the GLADE and the AllWISE catalogs to derive a reliable stellar mass estimation.

\begin{figure}
\begin{center}
\includegraphics[width=1.0\columnwidth]{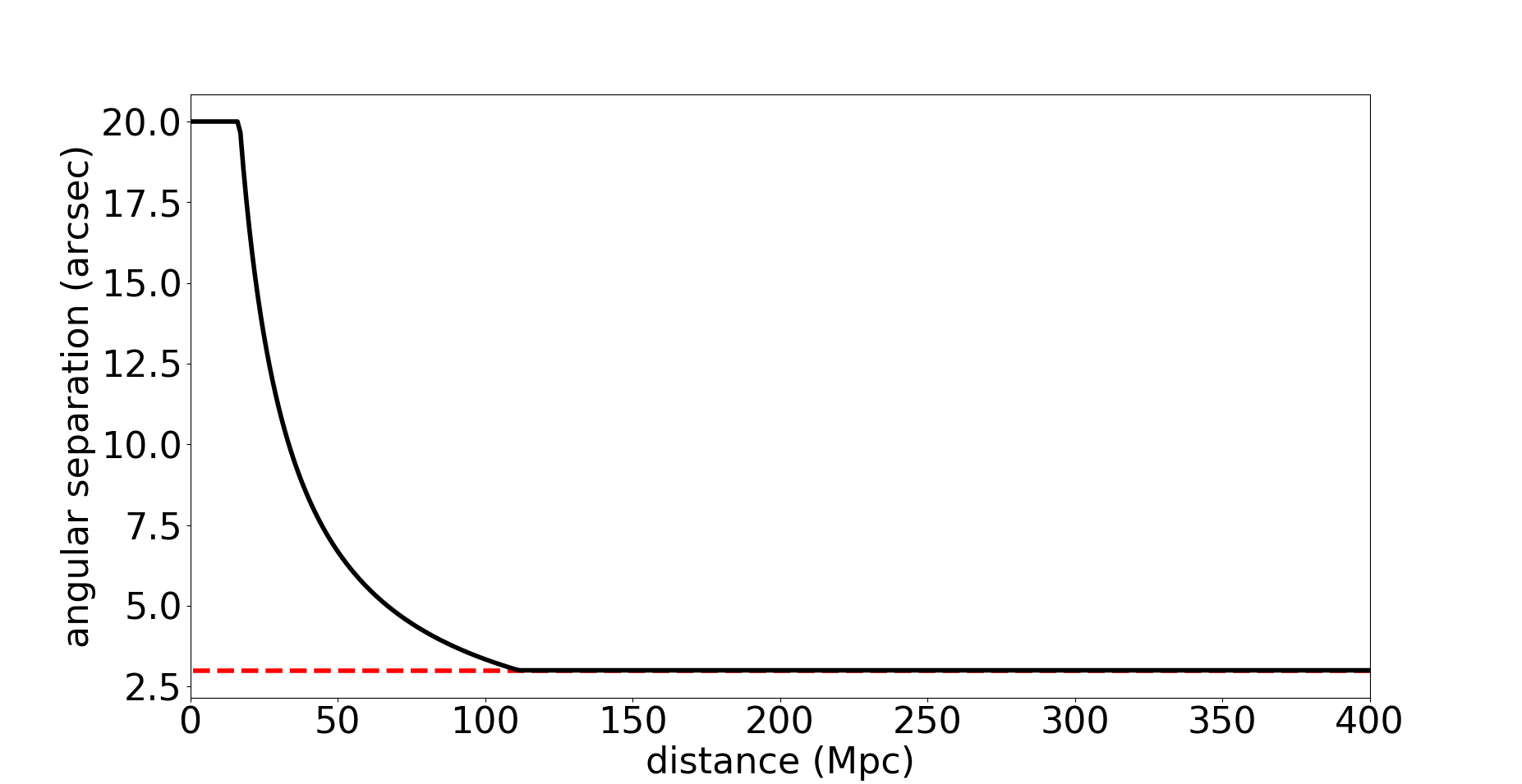}
\caption{Angular separation condition used for the crossmatch as a function of the distance. The red dashed line show the limit value of 3 arcseconds.}
\label{fig:queryangle}
\end{center}
\end{figure}

%We rely on the GLADE galaxy catalog to retrieve the galaxies compatible for a given GW skymap. In this catalog, the B, J, H and K band magnitudes are provided for some of the galaxies. Previous work \citep{LosC} used the B band magnitude included in the GLADE catalog as an indicator of the stellar mass. However, the B band is sensitive to the star formation history and can be strongly affected by dust extinction \citep{dustext2, FM07}. The near-infrared luminosity emitted by the old stellar population is fairly insensitive to dust extinction, and is thus considered as a reliable indicator of the total stellar mass of a galaxy \citep{Bruzual2003,Maraston2005}. In our work, we restrict the distance to 400Mpc as it is reasonably above the limiting sensibility distance of LIGO-Virgo for binary neutron star merger with O3 sensibility \citep{BNSrangecalc}.
%In the version 2.3 of the GLADE catalog, only $\sim$67\% of the galaxies up to 400Mpc have a K band magnitude. As showed by \citep{MLWISE12}, the WISE1 (3.4 $\mu$m) luminosity is a reliable indicator of the stellar mass, and the AllWISE catalog \citep{AllWISE} is all-sky. Therefore we perform a crossmatch with the AllWISE catalog to increase the number of galaxies with a near infrared magnitude to derive a reliable stellar mass estimation. \\
From WISE1 luminosities, \citep{MLWISE11,MLWISE12} showed that the stellar mass of galaxies can be reliably estimated with a constant mass to light ratio: $\Upsilon_{*}^{3.4\mu m} \sim 0.60 M_{\odot}/L_{\odot,3.4\mu m}$
where $M_{\odot}/L_{\odot,3.4\mu m}$ is the mass-to-light ratio in units of solar masses over the solar luminosity in the WISE 3.4$\mu$m band ($m_{\odot},3.4\mu m =3.24 mag; L_{\odot},3.4\mu m = 1.58 \times 10^{32} erg s^{-1} $; \citep{Jarrett2013}). This approach derives stellar mass with an error of 0.10 dex \citep{MLWISE11,MLWISE12}.
Although the value of $\Upsilon_{*}^{3.4\mu m}$ can vary from $\sim$0.5 to $\sim$0.65 in the literature, this results on small changes on the derived stellar mass and will not affect the ranking of the galaxies as it is a constant ratio.\\
We spatially crossmatched the GLADE catalog cut to 400Mpc with the AllWISE catalog using a radius crossmatch varying with the distance as seen in Figure \ref{fig:queryangle}. The radius is defined as $5\%$ the angular diameter of the milky way (0.0324 Mpc) and we impose a minimum and maximum of 3 and 20 arcseconds respectively. This strategy was chosen to optimize the match at low distance where galaxies can have very large angular size.
%We have spatially crossmatched ALLWISE to the galaxies in the GLADE catalog that are spatially coincident (within 2"\textcolor{red}{tbd}), and restricted the catalog to galaxies up to 400 Mpc (to cover the LIGO-Virgo sensitivity range for the O3 run \textcolor{red}{(repetition d'un peu plus haut)}).
A same AllWISE object can appear more than once in the preliminary crossmatched catalog, meaning that there is more than one GLADE galaxy around the given AllWISE object. For such case, knowing the angular resolution of the WISE telescope in the WISE1 band (6.1 arc seconds), we only kept the closest object if it is the only one in a radius of 6.1 arc seconds around the GLADE galaxy.
The presence of an AGN implies a significant emission in the infrared and near-infrared \citep{AGNIRemission1,AGNIRemission2,AGNIRemission3}, and thus biased our stellar mass estimation for such galaxies. We identify active galactic nucleus (AGN) from the resulting catalog using the mid-infrared color criterion $W1-W2 \ge 0.8$ as used in \citep{AGN1,AGN2}. This corresponds to 3346 galaxies, that are still present in the catalog but without stellar mass estimation. \\
We use the elliptical aperture photometry flux from ALLWISE catalog whenever available for the source, otherwise the profile fitting photometry is used. This ensures that these fluxes should encompass each galaxy full radial extent. At the end we obtain the stellar mass information for 743780 objects, knowing that GLADE catalog cut to 400 Mpc have 800986 galaxies we have a $\sim 93\%$ of match efficiency.\\

\begin{figure}
\begin{center}
\includegraphics[width=1.0\columnwidth]{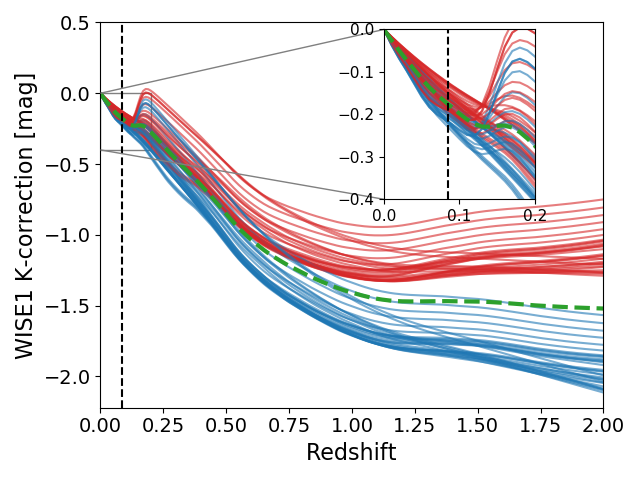}
\caption{K-correction for the 31 galaxy SED templates for the redshift range 0 to 2 with a zoom-in from 0 to 0.2. The blue solid lines are for the SED templates without dust attenuation, red solid lines are with a dust attenuation of $E (B-V)$=0.5 mag using the Calzetti law \citep{Calzetti00}. The green dashed line is the mean value at a given distance of the K-correction for all galaxies SED considered in this study. The black dashed line corresponds to 400Mpc.}
\label{fig:sed_kcorr}
\end{center}
\end{figure}

%\subsection{WISE1 k-correction}
%\label{wise1_kcorr}
Before the conversion of WISE1 magnitudes to stellar masses using the mass-to-light ratio provided by \citep{MLWISE11}, we apply a K-correction to correct for the distance. We use 31 SED templates covering all known type of galaxies, from red elliptical to blue star-forming galaxies \citep{Ilbert09}. We compute the K-correction both without dust attenuation and with $E(B-V)$=0.5 mag using the Calzetti attenuation law \citep{Calzetti00}. The K-correction is insensitive to the galaxy type and dust attenuation up to z=0.12 as seen in Figure \ref{fig:sed_kcorr}. Given the distance limitation of 400Mpc (z$\sim$0.085) for our catalog, we computed the K-correction at a given distance as the mean value for the 31 galaxies SED templates with and and without dust attenuation. At this range of distances, the effect is negligible, however this will become important for future catalogs that will need to be deeper to encompass the sensitivity improvement of LIGO-Virgo interferometers. We recall that all stellar masses used in this work were derived within this work using the same method to ensure homogeneity. Regarding the reliability of our stellar mass estimates, although we aimed at an homogeneous estimation, \citep{MLWISE11} showed that for low redshift galaxies the stellar mass estimation using a constant mas-to-light ration using the WISE1 luminosity is in good agreement with the one predicted from a more elaborate SED fitting technique.

\subsection{Completeness of the Mangrove catalog}
\label{section3}
In this section, we aim at quantifying the completeness of the Mangrove catalog in terms of stellar mass. The stellar mass function is well described by a Schechter function \citep{Schechter76}, $\Phi(M)$, which parametrises the number density of galaxies, $n_{galaxies}$, as a function of their stellar mass. In this work, we use the low-redshift galaxy stellar mass function derived from the Galaxy And Mass Assembly \citep{Wright17}, using a double Schechter function in logarithmic mass space defined as:
%\begin{equation}
%\label{eq:schechter_M}
%    N_{galaxies} = \Phi(M) ~ d log M
%\end{equation}
\begin{equation}
\label{eq:schechter_M}
\begin{split}
    n_{galaxies} & = \Phi(M) ~ d log M \\
    & = ln(10) ~e^{-10^{logM-logM^{*}}} \left[ \Phi^{*}_{1} . \left(10^{logM-logM^{*}}  \right)^{\alpha_{1}+1} \right. \\
    & \left.+ \Phi^{*}_{2} . \left(10^{logM-logM^{*}}  \right)^{\alpha_{2}+1} \right] ~ d log M
\end{split}
\end{equation}
where $logM^{*}$= 10.78, $\Phi^{*}_1$ = 2.93$\cdot 10^{-3} ~ h^3.Mpc^{-3}$, $\Phi^{*}_2$ = 0.63$\cdot 10^{-3} ~ h^3.Mpc^{-3}$, $\alpha_1$ = -0.62 and $\alpha_2$ = -1.5 \citep{Wright17}.\\
Following the approach of \citep{2016ApJ...820..136G, GLADEcatalog} regarding the luminosity function, we divided galaxies into 12 luminosity distance shells, with a width of 33.3 Mpc. For each shell, we construct histograms of WISE1 band derived stellar mass and integrate the double Schechter function for the same stellar mass bins in Figure \ref{fig:Schechter_M}. As the luminosity distance increases more and more low mass galaxies are missing. Regarding the double Schechter function derived from GAMA, the new catalog is fairly complete up to 33 Mpc in terms of stellar mass. In this work, we are interested in the more massive galaxies, so we computed the stellar mass, $M_{1/2}$, for which half of the stellar mass density is contributed by galaxies below and above this value. By computing $\int_{logM_{1/2}}^{13} logM \Phi(M) dlogM = 0.5 * \int_{7}^{13} logM \Phi(M) dlogM $, we find $log M_{1/2}$ = 10.674. For all the luminosity distance shells, the Mangrove catalog distribution follows the double Schechter function for $log M$ > $log M_{1/2}$ as seen in  Figure \ref{fig:Schechter_M}. Our method is prioritising the more massive galaxies, consequently the fairly completeness relative to the double Schechter function for galaxies at $log M$ > $log M_{1/2}$ minimises the lack of low mass galaxies as the distance increases. 
% To follow the approach of \citep{GLADEcatalog} where they fitted the Schechter function using the B band luminosity we divided galaxies into twelve luminosity distance boxes, with a width of $\Delta d_{L} = 16.7 Mpc$ for each one. We constructed (\textcolor{red}{see figure XXX}) histograms of WISE1 band luminosity for each boxes and compared it with the Schechter function for which the integration by $x = L_{W1}/L_{W1}^{*}$ is defined as:
%
%\begin{equation}
%\label{eq7}
%    N = \phi^{*}L_{W1}^{*} \Gamma(a+2,x_{1})
%\end{equation}

%Where following \citep{GLADEcatalog} we chosed $x_{1} = 0.626$ to focus on the brighter
%subset of galaxies. As in \citep{GLADEcatalog} we simply excluded galaxies for which we don't have the band of interest (here WISE1) from our completeness measurements. For us this represent a few number of galaxies as we have the WISE1 band for more than $96\%$ of the galaxies within 200Mpc.
\section{Validation on GW170817 event}
\label{section:GW170817}

At the time the event of the 17th august 2017 is the only gravitational wave event with a detected electromagnetic counterpart so our new method has to be tested on this event.\\
The GW170817 released distance is $40\pm 8$ Mpc, the $90\%$ skymap spans around $30 deg^{2}$ and as shown in the figure \ref{fig:gw170817galtar} according to our criteria (see Section \ref{section:gal_target} for the definition) there is 65 compatible galaxies within.
The table in Appendix \ref{appendixA} shows the results for this event on the selection of galaxies and their ranking according to the standard approach (Equation \ref{eq1}), according to \cite{LosC}, according to our method with product (Equation \ref{eq3}) and according to our method with addition (Equation \ref{eq4}). NGC 4993 host galaxy of GW170817 is ranked in $5^{th}$ position in the standard approach. The \citep{LosC} grade improved its rank to the $2^{nd}$ position and is ranked first with our method for both (Equation \ref{eq3}) and (Equation \ref{eq4}) expressions. These results show that if we had used our grade to monitor this event, there would have been a gain in the speed of observation of the host galaxy. An important thing to see in those results is that the new grades are behaving like expected i.e. galaxies with high stellar mass are prioritized compared to galaxies with small stellar mass (see Appendix \ref{appendixA} where the list of galaxies for the four methods are reported with their grade and stellar mass if used). We also stress that galaxies with a high 3D probability but without stellar mass information also appear in the list (see Appendix \ref{appendixA}), for instance galaxy WINGSJ125701.38-172325.2 is ranked in $9^{th}$ position, keeping such host candidate is a real improvement of the equation \ref{eq4} compared to equation \ref{eq3}.\\
\begin{figure}
\begin{center}
\includegraphics[width=1.0\columnwidth]{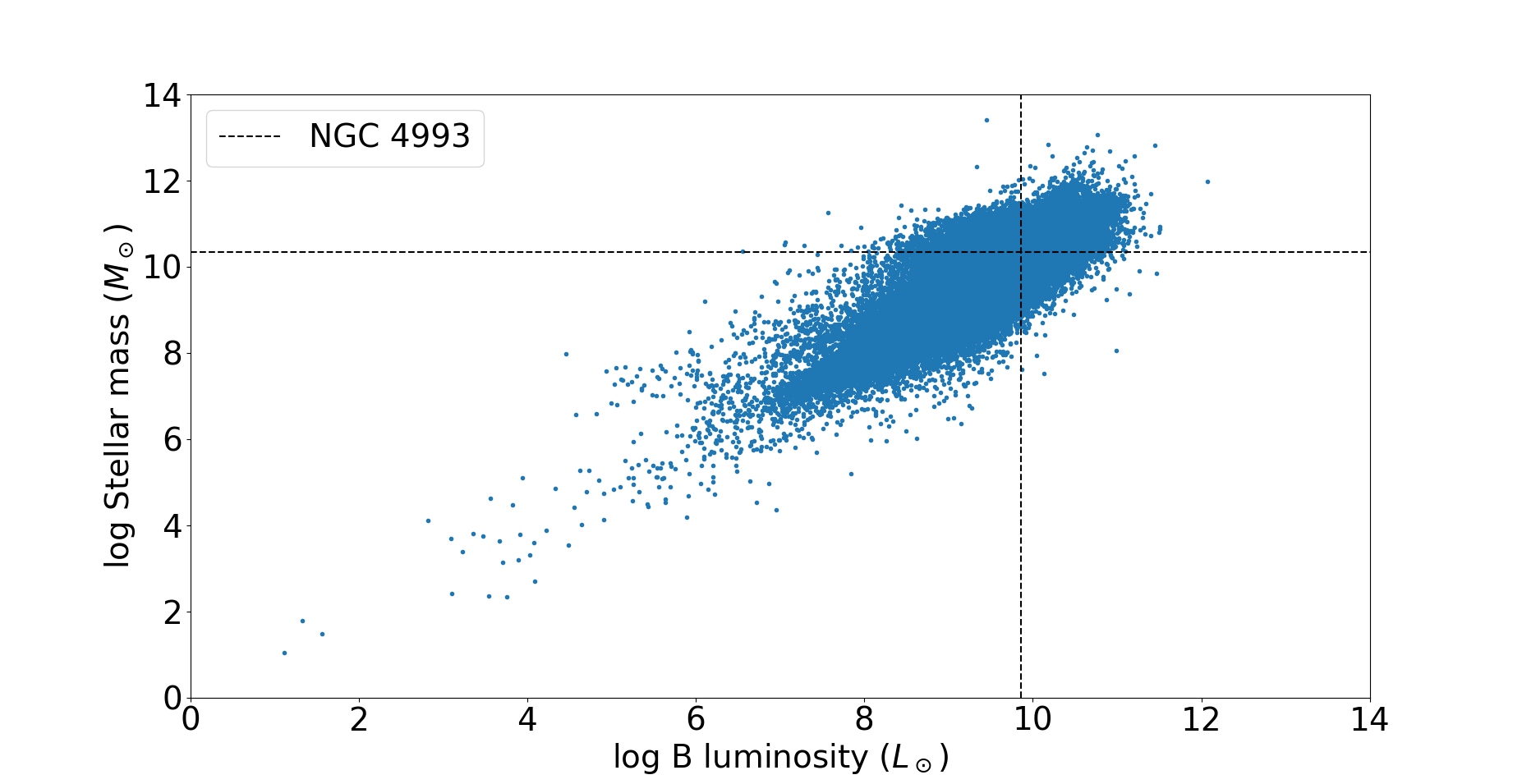}
\caption{B band luminosity provided by GLADE as a function of the stellar mass determined with the constant mass to light ratio using the W1 band. The crossing of the dashed lines shows the NGC 4993 (host of GW170817) position on this plot.}
\label{fig:lumB(stellarmass)}
\end{center}
\end{figure}
GW170817 was quite a lucky event in terms of distance of the source and good localisation thanks to data available for the three GW detectors. However, as the three inteferometers are not always in operating mode, this is very common to have a two inteferometers detection resulting in a poorer localisation \citep{LVCprospects}. In order to test our method on a larger skymap, i.e. a two inteferometers event, we choose to use the GW170817 skymap without Virgo data. The $90\%$ skymap spans $\sim 190 deg^{2}$, this is a good example of two GW detectors localisation for which we can be sure of the counterpart host. According to our criteria in Section \ref{section:gal_target}, there are 205 galaxies compatible with this skymap. The table in appendix \ref{appendixB} shows the resulting ranking for this skymap using the four presented grades. NGC 4993 host galaxy of the event is ranked in $27^{th}$ position in the standard approach. The \citep{LosC} grade ranks it at the $6^{th}$ position and our grade is even more successful by putting NGC 4993 in the $4^{th}$ position with both equations \ref{eq3} and \ref{eq4} expressions. Those results shows that the gain of our method for the follow up of gravitational waves event is even bigger in the case of wide skymap. This example of wider skymap allows us to see that sometime the \citep{LosC} grade and our grades behave differently and even oppositely. For example the galaxy PGC043966 is ranked in $37^{th}$ position with the standard grade and, when the \cite{LosC} grade upgrade its rank to the $30^{th}$, both of our grades (equations \ref{eq3}) and (\ref{eq4}) downgrade its rank to the $54^{th}$ and $49^{th}$ position respectively. We can also note that the \citep{LosC} grade ranked the NGC 4658 galaxy in first position due to is high B band luminosity but this galaxy is ranked in position 9 with our final grade definition. For these two examples the stellar mass estimation is not as high as the B band luminosity might suggest. Using the B band luminosity in those cases would have led to observe preferentially less massive galaxies.
We illustrate the difference in the behavior between our grade and the one from \citep{LosC} in Figure \ref{fig:lumB(stellarmass)}, where we plot the stellar mass estimation using WISE1 band as a function of the B band luminosity for the same galaxies. We see that for a given B band luminosity there is an important scatter in stellar mass spanning a few order of magnitudes. It means that the probability associated to a galaxy with respect to its stellar mass estimation can behave very differently for both methods. For example, for the host galaxy associated to GW170817, NGC 4993, we derived a stellar mass of $\sim 2.14 \times10^{10} M_\odot$ but galaxies with similar B band luminosity ($\sim 10^{10} L_{\odot}$ ) in the catalog span a stellar mass range from $\sim 3.8 \times10^{7} M_\odot$ to $\sim 1.0 \times10^{12} M_\odot$ which represents almost five order of magnitudes.
In addition to this different behavior, we also illustrate the flexibility of the grade defined in equation \ref{eq4} by noting that using the grade defined in equation \ref{eq3} would have led not to use $\sim7\%$ of the galaxies inside the 400Mpc of Mangrove catalog, and $\sim5\%$ of the galaxies when using the B band luminosity as in \citep{LosC}.
\onecolumn
\begingroup

\begin{figure}
\begin{center}
\includegraphics[width=1\columnwidth]{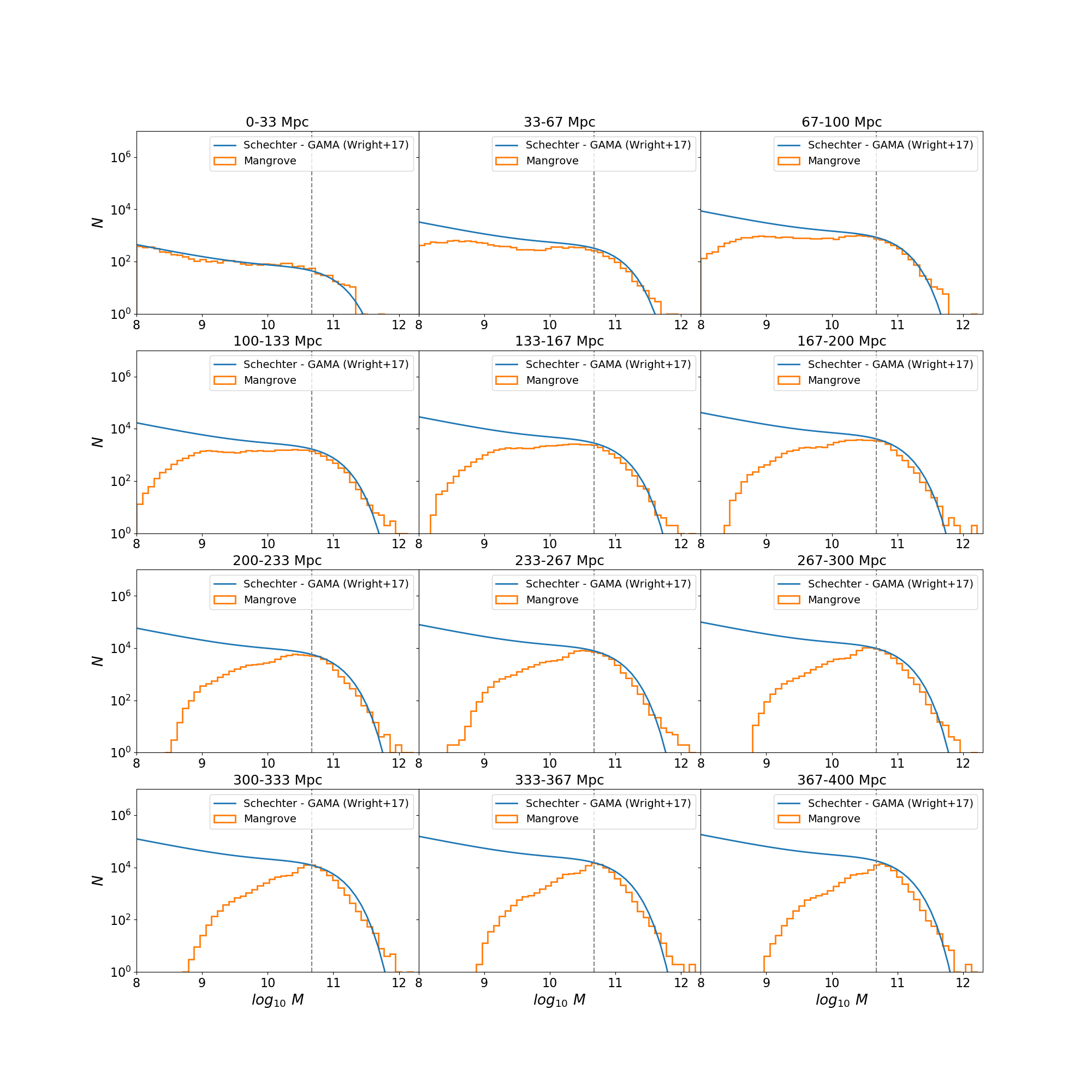}
\caption{Stellar mass histograms for the Mangrove catalog at different luminosity distance shells compared to the double Schechter function derived from GAMA \citep{Wright17} weighted by the volume of each shell. Each panel is divided in 50 bins of $log M$. The black dashed line represents the stellar mass, $log M_{1/2}$ for which half of the stellar mass density is contributed by galaxies at $log M$ > $log M_{1/2}$.}
\label{fig:Schechter_M}
\end{center}
\end{figure}
\twocolumn
\endgroup
%\textcolor{red}{discussion on grade value ?}

\section{Discussion and conclusion}
\label{section:conclusion}
\subsection{Utilisation of the galaxies for the tiling}

In this section we present a future development to use the galaxies for wide field of view telescopes. For the optimisation of wide FoV telescopes ($\gtrsim 1 deg^{2}$) follow up observations the standard approach consists in defining tiles over the sky and ranking them according to a given grade \citep{gwemopt,Ghosh2016}. A first version of the grade of a tile you can build is:
\begin{equation}
\label{eq:gradetile1}
\begin{split}
    Grade_{tile} = \sum_{pixel \in tile}{ P_{2D,pixel}}  
\end{split}
\end{equation}
where we sum up the 2D probability of the pixels $P_{2D,pixel}$ within the tile. When using a catalog of galaxies one can define a "galaxy weighted" grade for the tile using the grade of the galaxies:
\begin{equation}
\label{eq:gradetile2}
\begin{split}
    Grade_{tile} = \sum_{gal \in tile}{ Grade_{gal}}  
\end{split}
\end{equation}
where we sum up the grade of the galaxies within the tile. In this expression any galaxy grade can be used, such as the expression in equation \ref{eq4} that will optimise the chance to find the GW electromagnetic counterpart. The biggest issue of this approach is the catalog completeness, which makes this approach valid only below a distance threshold above which one has to switch back from grade definition of equation \ref{eq:gradetile2} to equation \ref{eq:gradetile1} in order to prevent using only galaxy information at a distance where your catalog is not complete enough.\\
We present a reformulation of the tile grade using our developments which allows to use galaxies catalog and a galaxy weighted grade at any distances. From Section \ref{section3}, we are able to define the mass completeness of the catalog, $C_{m1,m2}^{d1,d2}$, between distances $d_{1}$ and $d_{2}$ for a stellar mass range between $m_{1}$ and $m_{2}$ comparing to the double Schechter function:
\begin{equation}
\label{eq:defC}
\begin{split}
    C_{m1,m2}^{d1,d2} = \int_{m1}^{m2} M_{*} f_{Schecheter}(M_{*}) dM_{*}\\ - 
    \int_{m1}^{m2} M_{*} histogram(M_{*}) dM_{*}
\end{split}
\end{equation}
With this parameter defining the completeness, we can define the grade of a given tile by:
\begin{equation}
\label{eq:Ptile}
\begin{split}
    P_{tile} = \sum_{pixel \in tile}{\left[ \sum_{gal \in pixel}{C_{m1,m2}^{d1,d2}P_{tot,gal}}+(1-C_{m1,m2}^{d1,d2})P_{2D,pixel}
    \right]}
\end{split}
\end{equation}
where the first sum over all pixels inside the tile, the second sum over all galaxies falling in the 2D location of the pixel, $P_{tot,gal}$ is the grade of a given galaxy defined in Section \ref{subsection:Gradereformulation}, $P_{2D,pixel}$ is the 2D probability of the pixel, $d_{1}$ and $d_{2}$ are chosen as $\mu_{pixel} \mp \sigma_{pixel}$ respectively and $m_{1}$ and $m_{2}$ are fixed to $10^7$ and $10^{13}$ (stellar mass range validity of the Schechter function fitted by \cite{Wright17}). This expression removes any distance limitation for the use of galaxies weighted tiles. Note that with this expression we simply sump up the probability of the galaxies within a tile when the catalog is complete, whereas we sum up the 2D probability of all pixels within a tile when the catalog is not complete. 
% \begin{equation}
% \label{eq:Ptilelim1}
% \begin{split}
%     P_{tile} \xrightarrow{C_{m1,m2}^{d1,d2}\to 1}  \sum_{pixel \in tile}{\left[ \sum_{gal \in pixel}{P_{tot,gal}}\right]}  
% \end{split}
% \end{equation}

% \textcolor{red}{And:}

% \begin{equation}
% \label{eq:Ptilelim2}
% \begin{split}
%     P_{tile} \xrightarrow{C_{m1,m2}^{d1,d2}\to 0}  \sum_{pixel \in tile}{ P_{2D,pixel}}  
% \end{split}
% \end{equation}

\subsection{Conclusion}

The electromagnetic follow-up of gravitational wave events is very challenging, the poor localisation of the source provided by LIGO-Virgo forces telescopes around the world to observe large areas of the sky. As the electromagnetic counterpart is expected to decay rapidly in luminosity, an optimisation is required to perform a rapid and efficient follow-up of the skymap. Recent developments in both catalog of galaxies and galaxy targeting strategy already optimised significantly the follow-up of such event. Our work provides an efficient tool to upgrade in one hand a catalog of galaxies by adding the stellar mass information and on the other hand the galaxy targeting approach with a new expression of the grade using this stellar mass information to select and rank the galaxies. We crossmatched the GLADE and AllWISE catalogs to retrieve the WISE1 band luminosity to determine the stellar mass of $\sim$87\% of the galaxies inside the GLADE catalog up to 400Mpc. This catalog is complete in terms of stellar mass up to $\sim$33Mpc, and up to 200Mpc if we consider galaxies contributing to half of the stellar mass density. The new formulation of the grade presented in this work allows not only to use the 3D position of galaxies to select them but also their stellar mass. We tested and validated our grade on the GW170817 event by showing an improvement on the ranking of interesting galaxies, i.e. massive galaxies, where NGC4993 is ranked first. 
%A bigger sample of gravitational wave galaxy host population, not available at the time, will be useful to refine $\beta$ parameter of our grade definition.
This work plainly encourages further developments of the galaxy targeting strategy including other physical properties of the galaxies, for instance by focusing low SFR galaxies, but such development are slowed down by the poor number of information available in the publicly available galaxy catalogs. The Mangrove catalog is publicly available at \url{https://mangrove.lal.in2p3.fr}\footnote{The full catalog can be downloaded at \url{https://mangrove.lal.in2p3.fr/download_full.php}}, and this dedicated website automatically generates the list of galaxies, ranked by our new grade, compatible for each BNS event below 400Mpc, and observable from a given location on Earth. The improved grade presented in this work is implemented in the widely used \textit{gwemopt}\footnote{\url{https://github.com/mcoughlin/gwemopt}} python package \citep{gwemopt},
developed to optimize the efforts of electromagnetic follow-up of gravitational wave events.

\section*{Acknowledgements}

We acknowledge the Virtual data of labex P2IO for the supply of IT resources. This work made use of the NASA/IPAC Extragalactic Database (NED). We acknowledge the GRANDMA (Global Rapid Advanced Network Devoted to the Multi-messenger Addicts) collaboration for its technical support and assistance in the development of the mangrove web interface.

\bibliographystyle{mnras}
\bibliography{references}

\begin{thebibliography}{}
\makeatletter
\relax
\def\mn@urlcharsother{\let\do\@makeother \do\$\do\&\do\#\do\^\do\_\do\%\do\~}
\def\mn@doi{\begingroup\mn@urlcharsother \@ifnextchar [ {\mn@doi@}
  {\mn@doi@[]}}
\def\mn@doi@[#1]#2{\def\@tempa{#1}\ifx\@tempa\@empty \href
  {http://dx.doi.org/#2} {doi:#2}\else \href {http://dx.doi.org/#2} {#1}\fi
  \endgroup}
\def\mn@eprint#1#2{\mn@eprint@#1:#2::\@nil}
\def\mn@eprint@arXiv#1{\href {http://arxiv.org/abs/#1} {{\tt arXiv:#1}}}
\def\mn@eprint@dblp#1{\href {http://dblp.uni-trier.de/rec/bibtex/#1.xml}
  {dblp:#1}}
\def\mn@eprint@#1:#2:#3:#4\@nil{\def\@tempa {#1}\def\@tempb {#2}\def\@tempc
  {#3}\ifx \@tempc \@empty \let \@tempc \@tempb \let \@tempb \@tempa \fi \ifx
  \@tempb \@empty \def\@tempb {arXiv}\fi \@ifundefined
  {mn@eprint@\@tempb}{\@tempb:\@tempc}{\expandafter \expandafter \csname
  mn@eprint@\@tempb\endcsname \expandafter{\@tempc}}}

\bibitem[\protect\citeauthoryear{Abbott et~al.}{Abbott
  et~al.}{2017a}]{LSC_BNS_2017PhRvL}
Abbott B.~P.,  et~al., 2017a, \mn@doi [Phys. Rev. Lett.]
  {10.1103/PhysRevLett.119.161101}, 119, 161101

\bibitem[\protect\citeauthoryear{{Abbott} et~al.,}{{Abbott}
  et~al.}{2017b}]{gwtohubble1}
{Abbott} B.~P.,  et~al., 2017b, \mn@doi [\nat] {10.1038/nature24471}, \href
  {https://ui.adsabs.harvard.edu/abs/2017Natur.551...85A} {551, 85}

\bibitem[\protect\citeauthoryear{{Abbott} et~al.}{{Abbott}
  et~al.}{2017c}]{LSC_GW_GRB_2017ApJ}
{Abbott} B.~P.,  et~al., 2017c, \mn@doi [\apjl] {10.3847/2041-8213/aa920c},
  \href {http://adsabs.harvard.edu/abs/2017ApJ...848L..13A} {848, L13}

\bibitem[\protect\citeauthoryear{{Abbott} et~al.,}{{Abbott}
  et~al.}{2018a}]{BNSrange}
{Abbott} B.~P.,  et~al., 2018a, \mn@doi [Living Reviews in Relativity]
  {10.1007/s41114-018-0012-9}, \href
  {https://ui.adsabs.harvard.edu/abs/2018LRR....21....3A} {21, 3}

\bibitem[\protect\citeauthoryear{{Abbott} et~al.,}{{Abbott}
  et~al.}{2018b}]{LVCprospects}
{Abbott} B.~P.,  et~al., 2018b, \mn@doi [Living Reviews in Relativity]
  {10.1007/s41114-018-0012-9}, \href
  {https://ui.adsabs.harvard.edu/abs/2018LRR....21....3A} {21, 3}

\bibitem[\protect\citeauthoryear{{Abbott} et~al.,}{{Abbott}
  et~al.}{2018c}]{2018PhRvL.121p1101A}
{Abbott} B.~P.,  et~al., 2018c, \mn@doi [\prl]
  {10.1103/PhysRevLett.121.161101}, \href
  {https://ui.adsabs.harvard.edu/abs/2018PhRvL.121p1101A} {121, 161101}

\bibitem[\protect\citeauthoryear{{Antolini}, {Caiazzo}, {Dav{\'e}}  \&
  {Heyl}}{{Antolini} et~al.}{2017}]{Antolini2017}
{Antolini} E.,  {Caiazzo} I.,  {Dav{\'e}} R.,   {Heyl} J.~S.,  2017, \mn@doi
  [\mnras] {10.1093/mnras/stw3292}, \href
  {https://ui.adsabs.harvard.edu/abs/2017MNRAS.466.2212A} {466, 2212}

\bibitem[\protect\citeauthoryear{{Arcavi} et~al.,}{{Arcavi}
  et~al.}{2017}]{LosC}
{Arcavi} I.,  et~al., 2017, \mn@doi [\apjl] {10.3847/2041-8213/aa910f}, \href
  {https://ui.adsabs.harvard.edu/abs/2017ApJ...848L..33A} {848, L33}

\bibitem[\protect\citeauthoryear{{Artale}, {Mapelli}, {Giacobbo}, {Sabha},
  {Spera}, {Santoliquido}  \& {Bressan}}{{Artale}
  et~al.}{2019}]{2019MNRAS.487.1675A}
{Artale} M.~C.,  {Mapelli} M.,  {Giacobbo} N.,  {Sabha} N.~B.,  {Spera} M.,
  {Santoliquido} F.,   {Bressan} A.,  2019, \mn@doi [\mnras]
  {10.1093/mnras/stz1382}, \href
  {https://ui.adsabs.harvard.edu/abs/2019MNRAS.487.1675A} {487, 1675}

\bibitem[\protect\citeauthoryear{{Assef}, {Stern}, {Noirot}, {Jun}, {Cutri}  \&
  {Eisenhardt}}{{Assef} et~al.}{2018}]{AGN2}
{Assef} R.~J.,  {Stern} D.,  {Noirot} G.,  {Jun} H.~D.,  {Cutri} R.~M.,
  {Eisenhardt} P.~R.~M.,  2018, \mn@doi [\apjs] {10.3847/1538-4365/aaa00a},
  \href {https://ui.adsabs.harvard.edu/abs/2018ApJS..234...23A} {234, 23}

\bibitem[\protect\citeauthoryear{{Berger}}{{Berger}}{2014}]{Berger2014}
{Berger} E.,  2014, \mn@doi [\araa] {10.1146/annurev-astro-081913-035926},
  \href {https://ui.adsabs.harvard.edu/abs/2014ARA&A..52...43B} {52, 43}

\bibitem[\protect\citeauthoryear{{Blanchard} et~al.,}{{Blanchard}
  et~al.}{2017}]{Blanchard2017}
{Blanchard} P.~K.,  et~al., 2017, \mn@doi [\apjl] {10.3847/2041-8213/aa9055},
  \href {https://ui.adsabs.harvard.edu/abs/2017ApJ...848L..22B} {848, L22}

\bibitem[\protect\citeauthoryear{{Bruzual} \& {Charlot}}{{Bruzual} \&
  {Charlot}}{2003}]{Bruzual2003}
{Bruzual} G.,  {Charlot} S.,  2003, \mn@doi [\mnras]
  {10.1046/j.1365-8711.2003.06897.x}, \href
  {https://ui.adsabs.harvard.edu/abs/2003MNRAS.344.1000B} {344, 1000}

\bibitem[\protect\citeauthoryear{{Burtscher} et~al.,}{{Burtscher}
  et~al.}{2015}]{AGNIRemission2}
{Burtscher} L.,  et~al., 2015, \mn@doi [\aap] {10.1051/0004-6361/201525817},
  \href {https://ui.adsabs.harvard.edu/abs/2015A&A...578A..47B} {578, A47}

\bibitem[\protect\citeauthoryear{{Calzetti}, {Armus}, {Bohlin}, {Kinney},
  {Koornneef}  \& {Storchi-Bergmann}}{{Calzetti} et~al.}{2000}]{Calzetti00}
{Calzetti} D.,  {Armus} L.,  {Bohlin} R.~C.,  {Kinney} A.~L.,  {Koornneef} J.,
   {Storchi-Bergmann} T.,  2000, \mn@doi [\apj] {10.1086/308692}, \href
  {https://ui.adsabs.harvard.edu/abs/2000ApJ...533..682C} {533, 682}

\bibitem[\protect\citeauthoryear{{Cardelli}, {Clayton}  \& {Mathis}}{{Cardelli}
  et~al.}{1989}]{dustext2}
{Cardelli} J.~A.,  {Clayton} G.~C.,   {Mathis} J.~S.,  1989, \mn@doi [\apj]
  {10.1086/167900}, \href
  {https://ui.adsabs.harvard.edu/abs/1989ApJ...345..245C} {345, 245}

\bibitem[\protect\citeauthoryear{{Chen}, {Holz}, {Miller}, {Evans}, {Vitale}
  \& {Creighton}}{{Chen} et~al.}{2017}]{BNSrangecalc}
{Chen} H.-Y.,  {Holz} D.~E.,  {Miller} J.,  {Evans} M.,  {Vitale} S.,
  {Creighton} J.,  2017, arXiv e-prints, \href
  {https://ui.adsabs.harvard.edu/abs/2017arXiv170908079C} {p. arXiv:1709.08079}

\bibitem[\protect\citeauthoryear{{Cook} et~al.,}{{Cook}
  et~al.}{2019}]{CLUcatalog}
{Cook} D.~O.,  et~al., 2019, \mn@doi [\apj] {10.3847/1538-4357/ab2131}, \href
  {https://ui.adsabs.harvard.edu/abs/2019ApJ...880....7C} {880, 7}

\bibitem[\protect\citeauthoryear{{Coughlin} et~al.,}{{Coughlin}
  et~al.}{2018}]{gwemopt}
{Coughlin} M.~W.,  et~al., 2018, \mn@doi [\mnras] {10.1093/mnras/sty1066},
  \href {https://ui.adsabs.harvard.edu/abs/2018MNRAS.478..692C} {478, 692}

\bibitem[\protect\citeauthoryear{{Coughlin}, {Dietrich}, {Heinzel}, {Khetan},
  {Antier}, {Christensen}, {Coulter}  \& {Foley}}{{Coughlin}
  et~al.}{2019a}]{gwtohubble2}
{Coughlin} M.~W.,  {Dietrich} T.,  {Heinzel} J.,  {Khetan} N.,  {Antier} S.,
  {Christensen} N.,  {Coulter} D.~A.,   {Foley} R.~J.,  2019a, arXiv e-prints,
  \href {https://ui.adsabs.harvard.edu/abs/2019arXiv190800889C} {p.
  arXiv:1908.00889}

\bibitem[\protect\citeauthoryear{{Coughlin} et~al.,}{{Coughlin}
  et~al.}{2019b}]{Coughlin19_opt}
{Coughlin} M.~W.,  et~al., 2019b, arXiv e-prints, \href
  {https://ui.adsabs.harvard.edu/abs/2019arXiv190901244C} {p. arXiv:1909.01244}

\bibitem[\protect\citeauthoryear{{Cutri} \& {et al.}}{{Cutri} \& {et
  al.}}{2014}]{AllWISE}
{Cutri} R.~M.,  {et al.} 2014, VizieR Online Data Catalog, \href
  {https://ui.adsabs.harvard.edu/abs/2014yCat.2328....0C} {p. II/328}

\bibitem[\protect\citeauthoryear{{D{\'a}lya} et~al.,}{{D{\'a}lya}
  et~al.}{2018}]{GLADEcatalog}
{D{\'a}lya} G.,  et~al., 2018, \mn@doi [\mnras] {10.1093/mnras/sty1703}, \href
  {https://ui.adsabs.harvard.edu/abs/2018MNRAS.479.2374D} {479, 2374}

\bibitem[\protect\citeauthoryear{{Fitzpatrick} \& {Massa}}{{Fitzpatrick} \&
  {Massa}}{2007}]{FM07}
{Fitzpatrick} E.~L.,  {Massa} D.,  2007, \mn@doi [\apj] {10.1086/518158}, \href
  {https://ui.adsabs.harvard.edu/abs/2007ApJ...663..320F} {663, 320}

\bibitem[\protect\citeauthoryear{{Fong} et~al.,}{{Fong}
  et~al.}{2013}]{Fong2013}
{Fong} W.,  et~al., 2013, \mn@doi [\apj] {10.1088/0004-637X/769/1/56}, \href
  {https://ui.adsabs.harvard.edu/abs/2013ApJ...769...56F} {769, 56}

\bibitem[\protect\citeauthoryear{{Gehrels}, {Cannizzo}, {Kanner}, {Kasliwal},
  {Nissanke}  \& {Singer}}{{Gehrels} et~al.}{2016}]{2016ApJ...820..136G}
{Gehrels} N.,  {Cannizzo} J.~K.,  {Kanner} J.,  {Kasliwal} M.~M.,  {Nissanke}
  S.,   {Singer} L.~P.,  2016, \mn@doi [\apj] {10.3847/0004-637X/820/2/136},
  \href {https://ui.adsabs.harvard.edu/abs/2016ApJ...820..136G} {820, 136}

\bibitem[\protect\citeauthoryear{{Ghosh}, {Bloemen}, {Nelemans}, {Groot}  \&
  {Price}}{{Ghosh} et~al.}{2016a}]{2016A&A...592A..82G}
{Ghosh} S.,  {Bloemen} S.,  {Nelemans} G.,  {Groot} P.~J.,   {Price} L.~R.,
  2016a, \mn@doi [\aap] {10.1051/0004-6361/201527712}, \href
  {https://ui.adsabs.harvard.edu/abs/2016A&A...592A..82G} {592, A82}

\bibitem[\protect\citeauthoryear{{Ghosh}, {Bloemen}, {Nelemans}, {Groot}  \&
  {Price}}{{Ghosh} et~al.}{2016b}]{Ghosh2016}
{Ghosh} S.,  {Bloemen} S.,  {Nelemans} G.,  {Groot} P.~J.,   {Price} L.~R.,
  2016b, \mn@doi [\aap] {10.1051/0004-6361/201527712}, \href
  {https://ui.adsabs.harvard.edu/abs/2016A&A...592A..82G} {592, A82}

\bibitem[\protect\citeauthoryear{{Gill}, {Nathanail}  \& {Rezzolla}}{{Gill}
  et~al.}{2019}]{2019ApJ...876..139G}
{Gill} R.,  {Nathanail} A.,   {Rezzolla} L.,  2019, \mn@doi [\apj]
  {10.3847/1538-4357/ab16da}, \href
  {https://ui.adsabs.harvard.edu/abs/2019ApJ...876..139G} {876, 139}

\bibitem[\protect\citeauthoryear{{G{\'o}rski}, {Hivon}, {Banday}, {Wand elt},
  {Hansen}, {Reinecke}  \& {Bartelmann}}{{G{\'o}rski} et~al.}{2005}]{HEALPix}
{G{\'o}rski} K.~M.,  {Hivon} E.,  {Banday} A.~J.,  {Wand elt} B.~D.,  {Hansen}
  F.~K.,  {Reinecke} M.,   {Bartelmann} M.,  2005, \mn@doi [\apj]
  {10.1086/427976}, \href
  {https://ui.adsabs.harvard.edu/abs/2005ApJ...622..759G} {622, 759}

\bibitem[\protect\citeauthoryear{{Hajela} et~al.,}{{Hajela}
  et~al.}{2019}]{2019arXiv190906393H}
{Hajela} A.,  et~al., 2019, arXiv e-prints, \href
  {https://ui.adsabs.harvard.edu/abs/2019arXiv190906393H} {p. arXiv:1909.06393}

\bibitem[\protect\citeauthoryear{{Hjorth} et~al.,}{{Hjorth}
  et~al.}{2017}]{Hjorth2017}
{Hjorth} J.,  et~al., 2017, \mn@doi [\apjl] {10.3847/2041-8213/aa9110}, \href
  {https://ui.adsabs.harvard.edu/abs/2017ApJ...848L..31H} {848, L31}

\bibitem[\protect\citeauthoryear{{Hotokezaka}, {Nakar}, {Gottlieb}, {Nissanke},
  {Masuda}, {Hallinan}, {Mooley}  \& {Deller}}{{Hotokezaka}
  et~al.}{2019}]{gwtohubble3}
{Hotokezaka} K.,  {Nakar} E.,  {Gottlieb} O.,  {Nissanke} S.,  {Masuda} K.,
  {Hallinan} G.,  {Mooley} K.~P.,   {Deller} A.~T.,  2019, \mn@doi [Nature
  Astronomy] {10.1038/s41550-019-0820-1}, \href
  {https://ui.adsabs.harvard.edu/abs/2019NatAs...3..940H} {3, 940}

\bibitem[\protect\citeauthoryear{{Ilbert} et~al.,}{{Ilbert}
  et~al.}{2009}]{Ilbert09}
{Ilbert} O.,  et~al., 2009, \mn@doi [\apj] {10.1088/0004-637X/690/2/1236},
  \href {https://ui.adsabs.harvard.edu/abs/2009ApJ...690.1236I} {690, 1236}

\bibitem[\protect\citeauthoryear{{Im} et~al.,}{{Im} et~al.}{2017}]{NGC4993mass}
{Im} M.,  et~al., 2017, \mn@doi [\apjl] {10.3847/2041-8213/aa9367}, \href
  {https://ui.adsabs.harvard.edu/abs/2017ApJ...849L..16I} {849, L16}

\bibitem[\protect\citeauthoryear{{Jarrett} et~al.,}{{Jarrett}
  et~al.}{2013}]{Jarrett2013}
{Jarrett} T.~H.,  et~al., 2013, \mn@doi [\aj] {10.1088/0004-6256/145/1/6},
  \href {https://ui.adsabs.harvard.edu/abs/2013AJ....145....6J} {145, 6}

\bibitem[\protect\citeauthoryear{{Kettlety} et~al.,}{{Kettlety}
  et~al.}{2018}]{MLWISE11}
{Kettlety} T.,  et~al., 2018, \mn@doi [\mnras] {10.1093/mnras/stx2379}, \href
  {https://ui.adsabs.harvard.edu/abs/2018MNRAS.473..776K} {473, 776}

\bibitem[\protect\citeauthoryear{{Leibler} \& {Berger}}{{Leibler} \&
  {Berger}}{2010}]{Leibler2010}
{Leibler} C.~N.,  {Berger} E.,  2010, \mn@doi [\apj]
  {10.1088/0004-637X/725/1/1202}, \href
  {https://ui.adsabs.harvard.edu/abs/2010ApJ...725.1202L} {725, 1202}

\bibitem[\protect\citeauthoryear{{Levan} et~al.,}{{Levan}
  et~al.}{2017}]{Levan2017}
{Levan} A.~J.,  et~al., 2017, \mn@doi [\apjl] {10.3847/2041-8213/aa905f}, \href
  {https://ui.adsabs.harvard.edu/abs/2017ApJ...848L..28L} {848, L28}

\bibitem[\protect\citeauthoryear{{Mapelli}, {Giacobbo}, {Toffano}, {Ripamonti},
  {Bressan}, {Spera}  \& {Branchesi}}{{Mapelli}
  et~al.}{2018}]{2018MNRAS.481.5324M}
{Mapelli} M.,  {Giacobbo} N.,  {Toffano} M.,  {Ripamonti} E.,  {Bressan} A.,
  {Spera} M.,   {Branchesi} M.,  2018, \mn@doi [\mnras]
  {10.1093/mnras/sty2663}, \href
  {https://ui.adsabs.harvard.edu/abs/2018MNRAS.481.5324M} {481, 5324}

\bibitem[\protect\citeauthoryear{{Maraston}}{{Maraston}}{2005}]{Maraston2005}
{Maraston} C.,  2005, \mn@doi [\mnras] {10.1111/j.1365-2966.2005.09270.x},
  \href {https://ui.adsabs.harvard.edu/abs/2005MNRAS.362..799M} {362, 799}

\bibitem[\protect\citeauthoryear{{Metzger}}{{Metzger}}{2019}]{Metzgerkilo}
{Metzger} B.~D.,  2019, arXiv e-prints, \href
  {https://ui.adsabs.harvard.edu/abs/2019arXiv191001617M} {p. arXiv:1910.01617}

\bibitem[\protect\citeauthoryear{{Norris}, {Meidt}, {Van de Ven}, {Schinnerer},
  {Groves}  \& {Querejeta}}{{Norris} et~al.}{2014}]{MLWISE12}
{Norris} M.~A.,  {Meidt} S.,  {Van de Ven} G.,  {Schinnerer} E.,  {Groves} B.,
   {Querejeta} M.,  2014, \mn@doi [\apj] {10.1088/0004-637X/797/1/55}, \href
  {https://ui.adsabs.harvard.edu/abs/2014ApJ...797...55N} {797, 55}

\bibitem[\protect\citeauthoryear{{Pan} et~al.,}{{Pan} et~al.}{2017}]{Pan2017}
{Pan} Y.~C.,  et~al., 2017, \mn@doi [\apjl] {10.3847/2041-8213/aa9116}, \href
  {https://ui.adsabs.harvard.edu/abs/2017ApJ...848L..30P} {848, L30}

\bibitem[\protect\citeauthoryear{{Planck Collaboration}}{{Planck
  Collaboration}}{2016}]{planck15}
{Planck Collaboration} 2016, \mn@doi [\aap] {10.1051/0004-6361/201525830},
  \href {https://ui.adsabs.harvard.edu/abs/2016A&A...594A..13P} {594, A13}

\bibitem[\protect\citeauthoryear{{Rana} \& {Mooley}}{{Rana} \&
  {Mooley}}{2019}]{Rana2019}
{Rana} J.,  {Mooley} K.~P.,  2019, arXiv e-prints, \href
  {https://ui.adsabs.harvard.edu/abs/2019arXiv190407335R} {p. arXiv:1904.07335}

\bibitem[\protect\citeauthoryear{{Ruiz}, {Risaliti}, {Nardini}, {Panessa}  \&
  {Carrera}}{{Ruiz} et~al.}{2013}]{AGNIRemission1}
{Ruiz} A.,  {Risaliti} G.,  {Nardini} E.,  {Panessa} F.,   {Carrera} F.~J.,
  2013, \mn@doi [\aap] {10.1051/0004-6361/201015257}, \href
  {https://ui.adsabs.harvard.edu/abs/2013A&A...549A.125R} {549, A125}

\bibitem[\protect\citeauthoryear{{Salafia}, {Colpi}, {Branchesi},
  {Chassande-Mottin}, {Ghirlanda}, {Ghisellini}  \& {Vergani}}{{Salafia}
  et~al.}{2017}]{WaW}
{Salafia} O.~S.,  {Colpi} M.,  {Branchesi} M.,  {Chassande-Mottin} E.,
  {Ghirlanda} G.,  {Ghisellini} G.,   {Vergani} S.~D.,  2017, \mn@doi [\apj]
  {10.3847/1538-4357/aa850e}, \href
  {https://ui.adsabs.harvard.edu/abs/2017ApJ...846...62S} {846, 62}

\bibitem[\protect\citeauthoryear{{Sanders}}{{Sanders}}{1999}]{AGNIRemission3}
{Sanders} D.~B.,  1999, in {Terzian} Y.,  {Khachikian} E.,   {Weedman} D.,
  eds,  IAU Symposium Vol. 194, Activity in Galaxies and Related Phenomena.
  p.~25 (\mn@eprint {arXiv} {astro-ph/9903445})

\bibitem[\protect\citeauthoryear{{Schechter}}{{Schechter}}{1976}]{Schechter76}
{Schechter} P.,  1976, \mn@doi [\apj] {10.1086/154079}, \href
  {https://ui.adsabs.harvard.edu/abs/1976ApJ...203..297S} {203, 297}

\bibitem[\protect\citeauthoryear{{Singer} \& {Price}}{{Singer} \&
  {Price}}{2016}]{bayestar}
{Singer} L.~P.,  {Price} L.~R.,  2016, \mn@doi [\prd]
  {10.1103/PhysRevD.93.024013}, \href
  {https://ui.adsabs.harvard.edu/abs/2016PhRvD..93b4013S} {93, 024013}

\bibitem[\protect\citeauthoryear{{Singer} et~al.,}{{Singer}
  et~al.}{2016}]{GoingtheDistance}
{Singer} L.~P.,  et~al., 2016, \mn@doi [\apjl] {10.3847/2041-8205/829/1/L15},
  \href {https://ui.adsabs.harvard.edu/abs/2016ApJ...829L..15S} {829, L15}

\bibitem[\protect\citeauthoryear{{Stern} et~al.,}{{Stern} et~al.}{2012}]{AGN1}
{Stern} D.,  et~al., 2012, \mn@doi [\apj] {10.1088/0004-637X/753/1/30}, \href
  {https://ui.adsabs.harvard.edu/abs/2012ApJ...753...30S} {753, 30}

\bibitem[\protect\citeauthoryear{{Toffano}, {Mapelli}, {Giacobbo}, {Artale}  \&
  {Ghirlanda}}{{Toffano} et~al.}{2019}]{2019MNRAS.tmp.2085T}
{Toffano} M.,  {Mapelli} M.,  {Giacobbo} N.,  {Artale} M.~C.,   {Ghirlanda} G.,
   2019, \mn@doi [\mnras] {10.1093/mnras/stz2415}, \href
  {https://ui.adsabs.harvard.edu/abs/2019MNRAS.tmp.2085T} {p.~2085}

\bibitem[\protect\citeauthoryear{{Troja} et~al.,}{{Troja}
  et~al.}{2018}]{GRB150101B}
{Troja} E.,  et~al., 2018, \mn@doi [Nature Communications]
  {10.1038/s41467-018-06558-7}, \href
  {https://ui.adsabs.harvard.edu/abs/2018NatCo...9.4089T} {9, 4089}

\bibitem[\protect\citeauthoryear{{Veitch} et~al.,}{{Veitch}
  et~al.}{2015}]{LALInference}
{Veitch} J.,  et~al., 2015, \mn@doi [\prd] {10.1103/PhysRevD.91.042003}, \href
  {https://ui.adsabs.harvard.edu/abs/2015PhRvD..91d2003V} {91, 042003}

\bibitem[\protect\citeauthoryear{{Wright} et~al.,}{{Wright}
  et~al.}{2017}]{Wright17}
{Wright} A.~H.,  et~al., 2017, \mn@doi [\mnras] {10.1093/mnras/stx1149}, \href
  {https://ui.adsabs.harvard.edu/abs/2017MNRAS.470..283W} {470, 283}

\makeatother
\end{thebibliography}

%    \begingroup
%    \onecolumn
%    \begin{figure}
%    \begin{center}
%    \includegraphics[width=0.7\columnwidth]{combined2.png}
%    \caption
%    \label{fig:gw170817}
%    \end{center}
%    \end{figure}
%    \endgroup

%%%%%%%%%%%%%%%%% APPENDICES %%%%%%%%%%%%%%%%%%%%%
\clearpage
\appendix

\onecolumn
\section{}
\label{appendixA}
\small
\begin{longtable}{|c|cc|cc|cc|cc|c|c|}
\caption{Ranking of the galaxies compatible with the GW170817 skymap according to grades defined in equations (\ref{eq1}), (\ref{eq3}) and (\ref{eq4}), and the grade of \citep{LosC}.}
\label{table1}\\
% \begin{tabular}{|c|cc|cc|cc|cc|c|c|}
\cline{1-11}

 \multirow{2}{*}{Galaxy name} &\multicolumn{2}{c|}{(\ref{eq1})} & \multicolumn{2}{c|}{\citep{LosC}} & \multicolumn{2}{c|}{(\ref{eq3})} &  \multicolumn{2}{c|}{(\ref{eq4})}& \multirow{2}{*}{BLum $(L_{\odot})$} & \multirow{2}{*}{Stellar mass $(M_{\odot})$}\\
\cline{2-9}
        & Rank & $P_{loc}$ & Rank & $G_{tot}$ & Rank & $G_{tot}$ & Rank & $G_{tot}$ &  &  \\
        \hline
    \endfirsthead
\hline
 \multirow{2}{*}{Galaxy name} &\multicolumn{2}{c|}{(\ref{eq1})} & \multicolumn{2}{c|}{\citep{LosC}} & \multicolumn{2}{c|}{(\ref{eq3})} &  \multicolumn{2}{c|}{(\ref{eq4})}& \multirow{2}{*}{BLum $(L_{\odot})$} & \multirow{2}{*}{Stellar mass $(M_{\odot})$}\\
\cline{2-9}

        & Rank & $P_{loc}$ & Rank & $G_{tot}$ & Rank & $G_{tot}$ & Rank & $G_{tot}$ &  &  \\
\hline
    \endhead

    \hline \multicolumn{11}{|c|}{{Continued on next page}} \\ \hline
    \endfoot

    \endlastfoot

%\hline

ESO575-053 & 1 & 0.06 & 8 & 0.043 & 6 & 0.038 & 6 & 0.05 & 8.32e+35 & 9.675 \\
PGC803966 & 2 & 0.059 & 35 & 0.003 & 35 & 0.001 & 8 & 0.033 & 5.96e+34 & 7.835 \\
WINGSJ125701.38-172325.2 & 3 & 0.059 & 59 & $<0.001$ & -- & -- & 9 & 0.033 & 1.52e+33 & -- \\
ESO508-014 & 4 & 0.047 & 20 & 0.013 & 20 & 0.003 & 11 & 0.027 & 3.20e+35 & 8.605 \\
\textbf{NGC4993} & \textbf{5} & \textbf{0.046} & \textbf{2} & \textbf{0.111} & \textbf{1} & \textbf{0.22} & \textbf{1} & \textbf{0.123} & \textbf{2.79e+36} & \textbf{10.551} \\
PGC797164 & 6 & 0.046 & 18 & 0.014 & 17 & 0.005 & 10 & 0.028 & 3.60e+35 & 8.864 \\
ESO508-004 & 7 & 0.045 & 14 & 0.016 & 26 & 0.002 & 12 & 0.026 & 4.17e+35 & 8.416 \\
IC4197 & 8 & 0.04 & 1 & 0.137 & 2 & 0.195 & 2 & 0.109 & 3.96e+36 & 10.563 \\
ESO508-019 & 9 & 0.04 & 5 & 0.064 & 16 & 0.005 & 13 & 0.024 & 1.87e+36 & 9.004 \\
2MASS 13104593-2351566 & 10 & 0.038 & 13 & 0.027 & -- & -- & 14 & 0.021 & 8.14e+35 & -- \\
796755 & 11 & 0.037 & 44 & 0.001 & 38 & $<0.001$ & 15 & 0.02 & 3.43e+34 & 7.9711 \\
NGC4968 & 12 & 0.036 & 4 & 0.072 & -- & -- & 16 & 0.02 & 2.29e+36 & -- \\
6dFJ1309178-242256 & 13 & 0.034 & 33 & 0.003 & 36 & 0.001 & 17 & 0.019 & 1.05e+35 & 8.075 \\
ESO508-010 & 14 & 0.033 & 11 & 0.04 & 7 & 0.036 & 7 & 0.035 & 1.39e+36 & 9.914 \\
PGC169663 & 15 & 0.031 & 42 & 0.001 & 30 & 0.001 & 18 & 0.018 & 4.26e+34 & 8.171 \\
IC4180 & 16 & 0.027 & 6 & 0.063 & 5 & 0.105 & 4 & 0.062 & 2.73e+36 & 10.466 \\
PGC043966 & 17 & 0.024 & 15 & 0.016 & 29 & 0.001 & 20 & 0.014 & 7.77e+35 & 8.574 \\
PGC799951 & 18 & 0.021 & 36 & 0.003 & 37 & $<0.001$ & 22 & 0.012 & 1.66e+35 & 8.247 \\
WINGSJ125701.40-172325.3 & 19 & 0.021 & -- & -- & -- & -- & 23 & 0.011 & -- & -- \\
ESO508-015 & 20 & 0.02 & 19 & 0.013 & 42 & $<0.001$ & 24 & 0.011 & 8.01e+35 & 7.885 \\
ESO508-024 & 21 & 0.019 & 10 & 0.04 & -- & -- & 26 & 0.011 & 2.45e+36 & -- \\
ESO575-029 & 22 & 0.019 & 9 & 0.042 & 15 & 0.008 & 19 & 0.014 & 2.59e+36 & 9.49 \\
PGC169670 & 23 & 0.017 & 40 & 0.002 & 31 & 0.001 & 28 & 0.01 & 1.05e+35 & 8.428 \\
PGC772879 & 24 & 0.017 & 46 & 0.001 & 46 & $<0.001$ & 29 & 0.01 & 5.72e+34 & 7.862 \\
NGC4970 & 25 & 0.017 & 3 & 0.081 & 3 & 0.139 & 3 & 0.071 & 5.54e+36 & 10.791 \\
WINGSJ125701.40-172325.3 & 26 & 0.015 & -- & -- & -- & -- & 31 & 0.008 & -- & -- \\
NGC4830 & 27 & 0.011 & 7 & 0.048 & 4 & 0.11 & 5 & 0.055 & 5.16e+36 & 10.882 \\
PGC043664 & 28 & 0.01 & 16 & 0.015 & 14 & 0.01 & 27 & 0.01 & 1.71e+36 & 9.874 \\
ESO575-061 & 29 & 0.01 & 37 & 0.002 & 43 & $<0.001$ & 34 & 0.006 & 2.74e+35 & 8.156 \\
PGC044023 & 30 & 0.009 & 43 & 0.001 & 32 & 0.001 & 36 & 0.005 & 1.41e+35 & 8.693 \\
PGC044312 & 31 & 0.008 & 39 & 0.002 & 23 & 0.002 & 37 & 0.005 & 2.86e+35 & 9.321 \\
PGC044500 & 32 & 0.007 & 25 & 0.007 & 33 & 0.001 & 38 & 0.004 & 1.08e+36 & 8.787 \\
PGC044021 & 33 & 0.006 & 26 & 0.006 & 28 & 0.001 & 39 & 0.004 & 1.16e+36 & 9.188 \\
ESO508-033 & 34 & 0.006 & 24 & 0.007 & 12 & 0.012 & 30 & 0.009 & 1.29e+36 & 10.152 \\
WINGSJ125217.42-153054.2 & 35 & 0.006 & -- & -- & 40 & $<0.001$ & 40 & 0.003 & -- & 8.431 \\
ABELL\_1644:[D80]141 & 36 & 0.005 & -- & -- & -- & -- & 42 & 0.003 & -- & -- \\
ESO508-011 & 37 & 0.005 & 32 & 0.003 & 47 & $<0.001$ & 45 & 0.003 & 8.54e+35 & 8.302 \\
PGC044478 & 38 & 0.004 & 29 & 0.004 & 45 & $<0.001$ & 46 & 0.002 & 1.08e+36 & 8.479 \\
IC3799 & 39 & 0.004 & 12 & 0.028 & 11 & 0.012 & 33 & 0.008 & 8.50e+36 & 10.37 \\
PGC183552 & 40 & 0.004 & 38 & 0.002 & 24 & 0.002 & 43 & 0.003 & 7.35e+35 & 9.568 \\
ESO508-003 & 41 & 0.003 & 23 & 0.007 & 22 & 0.002 & 44 & 0.003 & 2.94e+36 & 9.754 \\
NGC4763 & 42 & 0.003 & 17 & 0.015 & 9 & 0.02 & 25 & 0.011 & 6.15e+36 & 10.729 \\
PGC044234 & 43 & 0.003 & 27 & 0.005 & 25 & 0.002 & 47 & 0.002 & 2.12e+36 & 9.679 \\
ESO508-007 & 44 & 0.003 & 49 & 0.001 & 53 & $<0.001$ & 50 & 0.001 & 2.82e+35 & 7.609 \\
PGC043908 & 45 & 0.003 & 30 & 0.004 & 18 & 0.003 & 41 & 0.003 & 1.68e+36 & 10.0 \\
ESO575-035 & 46 & 0.002 & 34 & 0.003 & 39 & $<0.001$ & 51 & 0.001 & 1.66e+36 & 9.018 \\
PGC043424 & 47 & 0.002 & 22 & 0.008 & 8 & 0.027 & 21 & 0.013 & 5.32e+36 & 11.051 \\
IC3831 & 48 & 0.002 & 28 & 0.004 & 13 & 0.011 & 35 & 0.006 & 3.07e+36 & 10.698 \\
PGC043505 & 49 & 0.002 & 55 & $<0.001$ & 44 & $<0.001$ & 53 & 0.001 & 1.66e+35 & 8.964 \\
NGC4756 & 50 & 0.001 & 21 & 0.009 & 10 & 0.016 & 32 & 0.008 & 6.89e+36 & 10.904 \\
PGC043344 & 51 & 0.001 & 47 & 0.001 & -- & -- & 54 & 0.001 & 5.20e+35 & -- \\
WINGSJ125252.62-152426.5 & 52 & 0.001 & -- & -- & -- & -- & 55 & 0.001 & -- & -- \\
ESO508-020 & 53 & 0.001 & 48 & 0.001 & -- & -- & 58 & 0.001 & 5.69e+35 & -- \\
PGC910856 & 54 & 0.001 & 57 & $<0.001$ & 49 & $<0.001$ & 56 & 0.001 & 1.06e+35 & 8.75 \\
PGC908166 & 55 & 0.001 & 56 & $<0.001$ & 48 & $<0.001$ & 57 & 0.001 & 1.71e+35 & 8.763 \\
PGC043823 & 56 & 0.001 & 45 & 0.001 & 27 & 0.001 & 52 & 0.001 & 1.03e+36 & 9.937 \\
PGC046026 & 57 & 0.001 & 41 & 0.001 & 21 & 0.002 & 49 & 0.002 & 1.72e+36 & 10.22 \\
NGC4724 & 58 & 0.001 & 31 & 0.004 & 19 & 0.003 & 48 & 0.002 & 4.58e+36 & 10.365 \\
PGC170205 & 59 & 0.001 & 50 & 0.001 & -- & -- & 60 & $<0.001$ & 1.01e+36 & -- \\
PGC043913 & 60 & 0.001 & 51 & $<0.001$ & 34 & 0.001 & 59 & 0.001 & 9.78e+35 & 9.87 \\
PGC937614 & 61 & $<0.001$ & 53 & $<0.001$ & 41 & $<0.001$ & 61 & $<0.001$ & 1.36e+36 & 9.79 \\
PGC943386 & 62 & $<0.001$ & 58 & $<0.001$ & 51 & $<0.001$ & 62 & $<0.001$ & 4.77e+35 & 8.724 \\
ESO575-041 & 63 & $<0.001$ & 54 & $<0.001$ & 50 & $<0.001$ & 63 & $<0.001$ & 1.20e+36 & 8.879 \\
2MASS 12492243-1321162 & 64 & $<0.001$ & 52 & $<0.001$ & -- & -- & 64 & $<0.001$ & 2.01e+36 & -- \\
PGC942354 & 65 & $<0.001$ & 60 & $<0.001$ & 52 & $<0.001$ & 65 & $<0.001$ & 3.23e+35 & 8.723 \\
\hline
% \end{tabular}
\end{longtable}

\section{}
\label{appendixB}
\small

\begin{longtable}{|c|cc|cc|cc|cc|c|c|}
\caption{Ranking of the galaxies compatible with the GW170817 without virgo data skymap according to grades defined in equations (\ref{eq1}), (\ref{eq3}) and (\ref{eq4}), and the grade of \citep{LosC}.}
\label{table2}\\
% \begin{tabular}{|c|cc|cc|cc|cc|c|c|}
\cline{1-11}

 \multirow{2}{*}{Galaxy name}  &\multicolumn{2}{c|}{(\ref{eq1})} & \multicolumn{2}{c|}{\citep{LosC}} & \multicolumn{2}{c|}{(\ref{eq3})} &  \multicolumn{2}{c|}{(\ref{eq4})}& \multirow{2}{*}{BLum $(L_{\odot})$} & \multirow{2}{*}{Stellar mass $(M_{\odot})$}\\
\cline{2-9}
        & Rank & $P_{loc}$ & Rank & $G_{tot}$ & Rank & $G_{tot}$ & Rank & $G_{tot}$ &  &  \\
        \hline
    \endfirsthead
\hline
\multirow{2}{*}{Galaxy name}  &\multicolumn{2}{c|}{(\ref{eq1})} & \multicolumn{2}{c|}{\citep{LosC}} & \multicolumn{2}{c|}{(\ref{eq3})} &  \multicolumn{2}{c|}{(\ref{eq4})}& \multirow{2}{*}{BLum $(L_{\odot})$} & \multirow{2}{*}{Stellar mass $(M_{\odot})$}\\
\cline{2-9}

        & Rank & $P_{loc}$ & Rank & $G_{tot}$ & Rank & $G_{tot}$ & Rank & $G_{tot}$ &  &  \\
\hline
    \endhead

    \hline \multicolumn{11}{|c|}{{Continued on next page}} \\ \hline
    \endfoot

    \endlastfoot

%\hline

ESO027-022 & 1 & 0.043 & 26 & 0.012 & 55 & 0.001 & 11 & 0.023 & 2.38e+35 & 7.958 \\
ESO027-003 & 2 & 0.029 & 7 & 0.034 & 42 & 0.001 & 15 & 0.016 & 9.85e+35 & 8.461 \\
NGC4348 & 3 & 0.028 & 2 & 0.058 & 1 & 0.145 & 1 & 0.083 & 1.73e+36 & 10.47 \\
PGC1108616 & 4 & 0.027 & 88 & 0.001 & 80 & $<0.001$ & 16 & 0.014 & 1.81e+34 & 7.497 \\
PGC3294456 & 5 & 0.027 & 101 & $<0.001$ & 92 & $<0.001$ & 17 & 0.014 & 1.14e+34 & 7.339 \\
PGC3293647 & 6 & 0.025 & 112 & $<0.001$ & 64 & $<0.001$ & 19 & 0.013 & 7.33e+33 & 7.795 \\
NGC4680 & 7 & 0.024 & 5 & 0.047 & 3 & 0.076 & 2 & 0.048 & 1.64e+36 & 10.252 \\
1143004 & 8 & 0.024 & 100 & $<0.001$ & -- & -- & 20 & 0.013 & 1.41e+34 & -- \\
229961 & 9 & 0.021 & 47 & 0.004 & 46 & 0.001 & 21 & 0.012 & 1.54e+35 & 8.5336 \\
NGC4663 & 10 & 0.019 & 14 & 0.025 & 10 & 0.046 & 6 & 0.032 & 1.13e+36 & 10.15 \\
2MASS 22302645-7941381 & 11 & 0.018 & 32 & 0.009 & 63 & $<0.001$ & 24 & 0.01 & 4.27e+35 & 8.07 \\
AGC229174 & 12 & 0.017 & -- & -- & -- & -- & 27 & 0.009 & -- & -- \\
GAMAJ121158.30+012934.6 & 13 & 0.016 & -- & -- & 102 & $<0.001$ & 29 & 0.008 & -- & 7.416 \\
NGC4658 & 14 & 0.015 & 1 & 0.102 & 8 & 0.047 & 8 & 0.03 & 5.59e+36 & 10.242 \\
PGC069012 & 15 & 0.015 & 21 & 0.015 & 35 & 0.002 & 26 & 0.009 & 8.28e+35 & 8.885 \\
ESO575-053 & 16 & 0.015 & 22 & 0.015 & 17 & 0.012 & 18 & 0.014 & 8.32e+35 & 9.675 \\
UGC07184 & 17 & 0.015 & 24 & 0.014 & 47 & 0.001 & 30 & 0.008 & 8.07e+35 & 8.523 \\
PGC803966 & 18 & 0.014 & 71 & 0.001 & 79 & $<0.001$ & 32 & 0.008 & 5.96e+34 & 7.835 \\
PGC1183373 & 19 & 0.014 & 115 & $<0.001$ & 125 & $<0.001$ & 35 & 0.007 & 1.01e+34 & 6.987 \\
WINGSJ125701.38-172325.2 & 20 & 0.014 & 155 & $<0.001$ & -- & -- & 36 & 0.007 & 1.52e+33 & -- \\
PGC797164 & 21 & 0.013 & 40 & 0.006 & 41 & 0.002 & 33 & 0.008 & 3.60e+35 & 8.864 \\
SDSSJ121210.92+025255.6 & 22 & 0.013 & 108 & $<0.001$ & 93 & $<0.001$ & 38 & 0.007 & 1.64e+34 & 7.653 \\
PGC1229057 & 23 & 0.013 & 83 & 0.001 & 70 & $<0.001$ & 39 & 0.007 & 4.22e+34 & 7.981 \\
GAMAJ122005.10+001556.4 & 24 & 0.012 & -- & -- & 86 & $<0.001$ & 40 & 0.007 & -- & 7.715 \\
PGC1066570 & 25 & 0.012 & 117 & $<0.001$ & 96 & $<0.001$ & 41 & 0.006 & 1.02e+34 & 7.65 \\
ESO508-019 & 26 & 0.012 & 12 & 0.026 & 34 & 0.002 & 37 & 0.007 & 1.87e+36 & 9.004 \\
\textbf{NGC4993} & \textbf{27} & \textbf{0.012} & \textbf{6} & \textbf{0.038} & \textbf{4} & \textbf{0.071} & \textbf{4} & \textbf{0.04} & \textbf{2.79e+36} & \textbf{10.551} \\
6dFJ1309178-242256 & 28 & 0.012 & 61 & 0.001 & 66 & $<0.001$ & 44 & 0.006 & 1.05e+35 & 8.075 \\
ESO508-004 & 29 & 0.012 & 39 & 0.006 & 59 & 0.001 & 43 & 0.006 & 4.17e+35 & 8.416 \\
ESO508-014 & 30 & 0.011 & 45 & 0.004 & 49 & 0.001 & 42 & 0.006 & 3.20e+35 & 8.605 \\
PGC1060528 & 31 & 0.011 & 89 & 0.001 & 73 & $<0.001$ & 46 & 0.006 & 4.31e+34 & 8.025 \\
PGC1193160 & 32 & 0.011 & 116 & $<0.001$ & 103 & $<0.001$ & 48 & 0.006 & 1.25e+34 & 7.563 \\
IC4197 & 33 & 0.011 & 4 & 0.051 & 5 & 0.069 & 5 & 0.038 & 3.96e+36 & 10.563 \\
796755 & 34 & 0.011 & 97 & $<0.001$ & 78 & $<0.001$ & 49 & 0.006 & 3.43e+34 & 7.9616 \\
PGC3294393 & 35 & 0.011 & 119 & $<0.001$ & 82 & $<0.001$ & 51 & 0.006 & 1.16e+34 & 7.885 \\
NGC4968 & 36 & 0.011 & 11 & 0.029 & -- & -- & 52 & 0.006 & 2.29e+36 & -- \\
PGC043966 & 37 & 0.01 & 30 & 0.01 & 54 & 0.001 & 47 & 0.006 & 7.77e+35 & 8.574 \\
ESO027-001 & 38 & 0.01 & 3 & 0.056 & 2 & 0.084 & 3 & 0.045 & 4.61e+36 & 10.671 \\
2MASS 00244271-7345157 & 39 & 0.01 & 27 & 0.011 & 38 & 0.002 & 45 & 0.006 & 9.43e+35 & 9.019 \\
AGC229200 & 40 & 0.01 & -- & -- & -- & -- & 54 & 0.005 & -- & -- \\
ESO027-008 & 41 & 0.01 & 8 & 0.033 & 6 & 0.054 & 7 & 0.031 & 2.85e+36 & 10.51 \\
3091844 & 42 & 0.01 & 103 & $<0.001$ & 89 & $<0.001$ & 55 & 0.005 & 2.95e+34 & 7.7983 \\
2MASS 13104593-2351566 & 43 & 0.009 & 31 & 0.009 & -- & -- & 56 & 0.005 & 8.14e+35 & -- \\
PGC3294258 & 44 & 0.009 & 129 & $<0.001$ & 123 & $<0.001$ & 57 & 0.005 & 8.62e+33 & 7.201 \\
SDSSJ120404.33+044847.2 & 45 & 0.009 & 102 & $<0.001$ & 90 & $<0.001$ & 61 & 0.005 & 3.39e+34 & 7.834 \\
PGC169663 & 46 & 0.009 & 98 & $<0.001$ & 69 & $<0.001$ & 60 & 0.005 & 4.26e+34 & 8.171 \\
PGC1166504 & 47 & 0.009 & 113 & $<0.001$ & 99 & $<0.001$ & 62 & 0.005 & 2.05e+34 & 7.713 \\
PGC772879 & 48 & 0.009 & 87 & 0.001 & 88 & $<0.001$ & 63 & 0.005 & 5.72e+34 & 7.862 \\
UGC07185 & 49 & 0.009 & 79 & 0.001 & 114 & $<0.001$ & 64 & 0.005 & 7.50e+34 & 7.457 \\
ESO508-010 & 50 & 0.008 & 25 & 0.014 & 18 & 0.012 & 23 & 0.01 & 1.39e+36 & 9.914 \\
ESO508-015 & 51 & 0.008 & 35 & 0.007 & 91 & $<0.001$ & 67 & 0.004 & 8.01e+35 & 7.885 \\
IC4180 & 52 & 0.008 & 15 & 0.024 & 12 & 0.038 & 12 & 0.022 & 2.73e+36 & 10.466 \\
NGC4123 & 53 & 0.008 & 16 & 0.02 & 19 & 0.01 & 28 & 0.009 & 2.18e+36 & 9.889 \\
PGC3294218 & 54 & 0.007 & 136 & $<0.001$ & 116 & $<0.001$ & 70 & 0.004 & 9.35e+33 & 7.516 \\
NGC4179 & 55 & 0.007 & 23 & 0.015 & 13 & 0.028 & 14 & 0.017 & 1.86e+36 & 10.377 \\
PGC044021 & 56 & 0.006 & 33 & 0.009 & 40 & 0.002 & 66 & 0.004 & 1.16e+36 & 9.188 \\
ESO027-021 & 57 & 0.006 & 28 & 0.01 & 20 & 0.01 & 31 & 0.008 & 1.41e+36 & 9.947 \\
SDSSJ120736.32+024143.3 & 58 & 0.006 & -- & -- & 61 & $<0.001$ & 72 & 0.003 & -- & 8.612 \\
PGC1233241 & 59 & 0.006 & 154 & $<0.001$ & 130 & $<0.001$ & 74 & 0.003 & 3.92e+33 & 7.074 \\
PGC799951 & 60 & 0.006 & 69 & 0.001 & 76 & $<0.001$ & 73 & 0.003 & 1.66e+35 & 8.247 \\
PGC039902 & 61 & 0.005 & 52 & 0.003 & 81 & $<0.001$ & 75 & 0.003 & 4.15e+35 & 8.183 \\
PGC037954 & 62 & 0.005 & 84 & 0.001 & 97 & $<0.001$ & 76 & 0.003 & 9.74e+34 & 7.971 \\
ESO508-024 & 63 & 0.005 & 20 & 0.015 & -- & -- & 77 & 0.003 & 2.45e+36 & -- \\
PGC043664 & 64 & 0.005 & 29 & 0.01 & 25 & 0.007 & 50 & 0.006 & 1.71e+36 & 9.874 \\
ESO575-029 & 65 & 0.005 & 19 & 0.015 & 32 & 0.003 & 68 & 0.004 & 2.59e+36 & 9.49 \\
NGC5967 & 66 & 0.005 & 13 & 0.026 & 9 & 0.047 & 10 & 0.024 & 4.44e+36 & 10.739 \\
J210518.19-824531.7 & 67 & 0.005 & -- & -- & 144 & $<0.001$ & 79 & 0.003 & -- & 6.727 \\
NGC4970 & 68 & 0.005 & 9 & 0.032 & 7 & 0.052 & 9 & 0.027 & 5.54e+36 & 10.791 \\
UGC07332 & 69 & 0.005 & 68 & 0.001 & 134 & $<0.001$ & 80 & 0.003 & 2.12e+35 & 7.079 \\
PGC2801913 & 70 & 0.005 & -- & -- & 27 & 0.005 & 59 & 0.005 & -- & 9.819 \\
NGC4116 & 71 & 0.005 & 38 & 0.006 & 33 & 0.002 & 71 & 0.003 & 1.14e+36 & 9.452 \\
PGC169670 & 72 & 0.004 & 91 & 0.001 & 72 & $<0.001$ & 81 & 0.002 & 1.05e+35 & 8.428 \\
SDSSJ121518.94+025538.2 & 73 & 0.004 & 145 & $<0.001$ & 129 & $<0.001$ & 82 & 0.002 & 7.72e+33 & 7.246 \\
UGC07178 & 74 & 0.004 & 72 & 0.001 & 132 & $<0.001$ & 83 & 0.002 & 1.95e+35 & 7.171 \\
WINGSJ125701.40-172325.3 & 75 & 0.004 & -- & -- & -- & -- & 85 & 0.002 & -- & -- \\
NGC5967A & 76 & 0.004 & 36 & 0.007 & -- & -- & 86 & 0.002 & 1.44e+36 & -- \\
UGC07396 & 77 & 0.004 & 42 & 0.005 & 67 & $<0.001$ & 84 & 0.002 & 1.11e+36 & 8.532 \\
2MASS 15465869-7547149 & 78 & 0.004 & 49 & 0.003 & -- & -- & 87 & 0.002 & 6.51e+35 & -- \\
3293713 & 79 & 0.003 & 152 & $<0.001$ & 126 & $<0.001$ & 90 & 0.002 & 7.70e+33 & 7.506 \\
ESO508-033 & 80 & 0.003 & 44 & 0.005 & 23 & 0.008 & 53 & 0.005 & 1.29e+36 & 10.152 \\
IC3799 & 81 & 0.003 & 10 & 0.031 & 16 & 0.013 & 34 & 0.008 & 8.50e+36 & 10.37 \\
3294175 & 82 & 0.003 & 164 & $<0.001$ & 153 & $<0.001$ & 93 & 0.002 & 2.11e+33 & 6.679 \\
NGC4830 & 83 & 0.003 & 17 & 0.018 & 11 & 0.04 & 13 & 0.02 & 5.16e+36 & 10.882 \\
GAMAJ121759.98+002558.1 & 84 & 0.003 & -- & -- & 154 & $<0.001$ & 94 & 0.002 & -- & 6.522 \\
WINGSJ125701.40-172325.3 & 85 & 0.003 & -- & -- & -- & -- & 95 & 0.002 & -- & -- \\
PGC135791 & 86 & 0.003 & 123 & $<0.001$ & 142 & $<0.001$ & 98 & 0.001 & 3.90e+34 & 7.022 \\
WINGSJ125217.42-153054.2 & 87 & 0.003 & -- & -- & 84 & $<0.001$ & 96 & 0.002 & -- & 8.431 \\
PGC043344 & 88 & 0.003 & 57 & 0.002 & -- & -- & 99 & 0.001 & 5.20e+35 & -- \\
PGC3271002 & 89 & 0.003 & 167 & $<0.001$ & 150 & $<0.001$ & 100 & 0.001 & 1.77e+33 & 6.814 \\
ESO042-007 & 90 & 0.003 & 34 & 0.008 & -- & -- & 101 & 0.001 & 2.64e+36 & -- \\
SDSSJ114850.14+102655.9 & 91 & 0.003 & -- & -- & 101 & $<0.001$ & 103 & 0.001 & -- & 8.229 \\
NGC3976 & 92 & 0.002 & 18 & 0.018 & 15 & 0.018 & 25 & 0.01 & 5.96e+36 & 10.613 \\
PGC037301 & 93 & 0.002 & 106 & $<0.001$ & 83 & $<0.001$ & 104 & 0.001 & 9.44e+34 & 8.488 \\
ESO575-061 & 94 & 0.002 & 77 & 0.001 & 107 & $<0.001$ & 105 & 0.001 & 2.74e+35 & 8.156 \\
PGC044023 & 95 & 0.002 & 99 & $<0.001$ & 71 & $<0.001$ & 102 & 0.001 & 1.41e+35 & 8.693 \\
PGC3122921 & 96 & 0.002 & 165 & $<0.001$ & 152 & $<0.001$ & 107 & 0.001 & 2.58e+33 & 6.807 \\
ESO508-035 & 97 & 0.002 & 105 & $<0.001$ & 117 & $<0.001$ & 108 & 0.001 & 1.01e+35 & 7.917 \\
GAMAJ121732.70+002646.3 & 98 & 0.002 & -- & -- & 109 & $<0.001$ & 109 & 0.001 & -- & 8.13 \\
ESO068-002 & 99 & 0.002 & 67 & 0.001 & 105 & $<0.001$ & 110 & 0.001 & 4.94e+35 & 8.221 \\
PGC044312 & 100 & 0.002 & 80 & 0.001 & 51 & 0.001 & 97 & 0.001 & 2.86e+35 & 9.321 \\
PGC1070576 & 101 & 0.002 & 94 & $<0.001$ & 108 & $<0.001$ & 112 & 0.001 & 2.11e+35 & 8.248 \\
6dFJ1258120-210246 & 102 & 0.002 & 95 & $<0.001$ & 95 & $<0.001$ & 114 & 0.001 & 2.18e+35 & 8.462 \\
PGC044500 & 103 & 0.002 & 53 & 0.002 & 74 & $<0.001$ & 113 & 0.001 & 1.08e+36 & 8.787 \\
SDSSJ115551.83+064354.9 & 104 & 0.002 & 135 & $<0.001$ & 122 & $<0.001$ & 118 & 0.001 & 3.90e+34 & 7.949 \\
ESO508-007 & 105 & 0.002 & 92 & 0.001 & 131 & $<0.001$ & 120 & 0.001 & 2.82e+35 & 7.609 \\
AGC215716 & 106 & 0.002 & 137 & $<0.001$ & 128 & $<0.001$ & 121 & 0.001 & 3.72e+34 & 7.743 \\
PGC039799 & 107 & 0.002 & 118 & $<0.001$ & 151 & $<0.001$ & 123 & 0.001 & 7.89e+34 & 7.024 \\
ABELL\_1644:[D80]141 & 108 & 0.002 & -- & -- & -- & -- & 124 & 0.001 & -- & -- \\
PGC037490 & 109 & 0.002 & 104 & $<0.001$ & 100 & $<0.001$ & 122 & 0.001 & 1.79e+35 & 8.473 \\
HIPASSJ1255-15 & 110 & 0.001 & -- & -- & -- & -- & 127 & 0.001 & -- & -- \\
ESO508-011 & 111 & 0.001 & 63 & 0.001 & 110 & $<0.001$ & 126 & 0.001 & 8.54e+35 & 8.302 \\
PGC044478 & 112 & 0.001 & 59 & 0.002 & 106 & $<0.001$ & 129 & 0.001 & 1.08e+36 & 8.479 \\
PGC183552 & 113 & 0.001 & 73 & 0.001 & 52 & 0.001 & 117 & 0.001 & 7.35e+35 & 9.568 \\
PGC170205 & 114 & 0.001 & 65 & 0.001 & -- & -- & 132 & 0.001 & 1.01e+36 & -- \\
IC3831 & 115 & 0.001 & 46 & 0.004 & 21 & 0.009 & 58 & 0.005 & 3.07e+36 & 10.698 \\
PGC043424 & 116 & 0.001 & 37 & 0.007 & 14 & 0.02 & 22 & 0.01 & 5.32e+36 & 11.051 \\
PGC1031551 & 117 & 0.001 & 163 & $<0.001$ & 159 & $<0.001$ & 134 & $<0.001$ & 7.36e+33 & 6.862 \\
PGC3294523 & 118 & 0.001 & 172 & $<0.001$ & 164 & $<0.001$ & 136 & $<0.001$ & 1.86e+33 & 6.498 \\
ESO508-003 & 119 & 0.001 & 51 & 0.003 & 50 & 0.001 & 125 & 0.001 & 2.94e+36 & 9.754 \\
PGC720745 & 120 & 0.001 & 153 & $<0.001$ & 140 & $<0.001$ & 138 & $<0.001$ & 3.11e+34 & 7.62 \\
IC0874 & 121 & 0.001 & 75 & 0.001 & 43 & 0.001 & 111 & 0.001 & 8.89e+35 & 10.016 \\
PGC758254 & 122 & 0.001 & 156 & $<0.001$ & 146 & $<0.001$ & 139 & $<0.001$ & 2.69e+34 & 7.49 \\
NGC4763 & 123 & 0.001 & 41 & 0.006 & 24 & 0.007 & 69 & 0.004 & 6.15e+36 & 10.729 \\
PGC043908 & 124 & 0.001 & 60 & 0.001 & 45 & 0.001 & 116 & 0.001 & 1.68e+36 & 10.0 \\
PGC3293619 & 125 & 0.001 & 180 & $<0.001$ & 162 & $<0.001$ & 141 & $<0.001$ & 6.67e+32 & 6.631 \\
PGC046026 & 126 & 0.001 & 64 & 0.001 & 36 & 0.002 & 106 & 0.001 & 1.72e+36 & 10.22 \\
NGC4724 & 127 & 0.001 & 48 & 0.004 & 31 & 0.003 & 92 & 0.002 & 4.58e+36 & 10.365 \\
PGC685308 & 128 & 0.001 & 130 & $<0.001$ & 104 & $<0.001$ & 142 & $<0.001$ & 1.21e+35 & 8.765 \\
PGC044234 & 129 & 0.001 & 58 & 0.002 & 57 & 0.001 & 131 & 0.001 & 2.12e+36 & 9.679 \\
PGC091191 & 130 & 0.001 & 168 & $<0.001$ & 163 & $<0.001$ & 145 & $<0.001$ & 5.86e+33 & 6.666 \\
NGC4756 & 131 & 0.001 & 43 & 0.005 & 22 & 0.008 & 65 & 0.004 & 6.89e+36 & 10.904 \\
WINGSJ125252.62-152426.5 & 132 & 0.001 & -- & -- & -- & -- & 148 & $<0.001$ & -- & -- \\
PGC3291384 & 133 & 0.001 & 179 & $<0.001$ & 170 & $<0.001$ & 149 & $<0.001$ & 8.83e+32 & 6.295 \\
135794 & 134 & 0.001 & 174 & $<0.001$ & 166 & $<0.001$ & 150 & $<0.001$ & 2.31e+33 & 6.473 \\
2MASXJ13242754-3025548 & 135 & 0.001 & 128 & $<0.001$ & -- & -- & 151 & $<0.001$ & 1.50e+35 & -- \\
PGC043505 & 136 & 0.001 & 125 & $<0.001$ & 98 & $<0.001$ & 147 & $<0.001$ & 1.66e+35 & 8.964 \\
ESO575-035 & 137 & 0.001 & 70 & 0.001 & 94 & $<0.001$ & 146 & $<0.001$ & 1.66e+36 & 9.018 \\
PGC043823 & 138 & 0.001 & 81 & 0.001 & 48 & 0.001 & 130 & 0.001 & 1.03e+36 & 9.937 \\
NGC4504 & 139 & 0.001 & 86 & 0.001 & 68 & $<0.001$ & 140 & $<0.001$ & 9.58e+35 & 9.402 \\
PGC3294233 & 140 & 0.001 & 178 & $<0.001$ & -- & -- & 152 & $<0.001$ & 9.60e+32 & -- \\
2MASXJ12490814-1124354 & 141 & 0.001 & 131 & $<0.001$ & -- & -- & 154 & $<0.001$ & 1.50e+35 & -- \\
PGC104686 & 142 & 0.001 & 148 & $<0.001$ & -- & -- & 155 & $<0.001$ & 5.98e+34 & -- \\
NGC5114 & 143 & 0.001 & 50 & 0.003 & 26 & 0.005 & 78 & 0.003 & 4.65e+36 & 10.774 \\
ESO508-036 & 144 & $<0.001$ & 146 & $<0.001$ & 138 & $<0.001$ & 156 & $<0.001$ & 6.54e+34 & 7.924 \\
PGC910856 & 145 & $<0.001$ & 139 & $<0.001$ & 111 & $<0.001$ & 153 & $<0.001$ & 1.06e+35 & 8.75 \\
PGC041725 & 146 & $<0.001$ & 133 & $<0.001$ & 157 & $<0.001$ & 158 & $<0.001$ & 1.60e+35 & 7.166 \\
NGC4487 & 147 & $<0.001$ & 85 & 0.001 & 65 & $<0.001$ & 144 & $<0.001$ & 1.14e+36 & 9.493 \\
PGC908166 & 148 & $<0.001$ & 134 & $<0.001$ & 113 & $<0.001$ & 159 & $<0.001$ & 1.71e+35 & 8.763 \\
PGC043913 & 149 & $<0.001$ & 96 & $<0.001$ & 58 & 0.001 & 135 & $<0.001$ & 9.78e+35 & 9.87 \\
PGC3097711 & 150 & $<0.001$ & 181 & $<0.001$ & 173 & $<0.001$ & 161 & $<0.001$ & 1.14e+33 & 6.106 \\
PGC141593 & 151 & $<0.001$ & 126 & $<0.001$ & 112 & $<0.001$ & 160 & $<0.001$ & 2.23e+35 & 8.8 \\
PGC3097710 & 152 & $<0.001$ & 182 & $<0.001$ & -- & -- & 164 & $<0.001$ & 1.14e+33 & -- \\
PGC705472 & 153 & $<0.001$ & 144 & $<0.001$ & 156 & $<0.001$ & 163 & $<0.001$ & 9.78e+34 & 7.363 \\
PGC937614 & 154 & $<0.001$ & 90 & 0.001 & 62 & $<0.001$ & 143 & $<0.001$ & 1.36e+36 & 9.79 \\
ESO508-020 & 155 & $<0.001$ & 110 & $<0.001$ & -- & -- & 165 & $<0.001$ & 5.69e+35 & -- \\
NGC5061 & 156 & $<0.001$ & 54 & 0.002 & 29 & 0.004 & 89 & 0.002 & 4.95e+36 & 10.805 \\
PGC135798 & 157 & $<0.001$ & 171 & $<0.001$ & 167 & $<0.001$ & 168 & $<0.001$ & 4.70e+33 & 6.665 \\
PGC042120 & 158 & $<0.001$ & 173 & $<0.001$ & 169 & $<0.001$ & 169 & $<0.001$ & 3.98e+33 & 6.529 \\
ESO444-026 & 159 & $<0.001$ & 56 & 0.002 & -- & -- & 170 & $<0.001$ & 4.64e+36 & -- \\
J132249.66-300651.8 & 160 & $<0.001$ & -- & -- & 119 & $<0.001$ & 166 & $<0.001$ & -- & 8.706 \\
PGC943386 & 161 & $<0.001$ & 114 & $<0.001$ & 118 & $<0.001$ & 167 & $<0.001$ & 4.77e+35 & 8.724 \\
PGC3097712 & 162 & $<0.001$ & 177 & $<0.001$ & 174 & $<0.001$ & 172 & $<0.001$ & 1.80e+33 & 6.13 \\
PGC3294387 & 163 & $<0.001$ & 184 & $<0.001$ & 172 & $<0.001$ & 171 & $<0.001$ & 4.59e+32 & 6.383 \\
PGC3268622 & 164 & $<0.001$ & 183 & $<0.001$ & 160 & $<0.001$ & 173 & $<0.001$ & 5.02e+32 & 7.236 \\
PGC3097709 & 165 & $<0.001$ & 176 & $<0.001$ & 171 & $<0.001$ & 174 & $<0.001$ & 2.38e+33 & 6.406 \\
PGC046803 & 166 & $<0.001$ & 93 & $<0.001$ & 53 & 0.001 & 133 & $<0.001$ & 1.43e+36 & 10.154 \\
PGC104868 & 167 & $<0.001$ & 158 & $<0.001$ & 148 & $<0.001$ & 175 & $<0.001$ & 4.53e+34 & 7.848 \\
ESO444-021 & 168 & $<0.001$ & 66 & 0.001 & 87 & $<0.001$ & 162 & $<0.001$ & 4.08e+36 & 9.381 \\
WINGSJ132507.84-315046.2 & 169 & $<0.001$ & -- & -- & -- & -- & 177 & $<0.001$ & -- & -- \\
NGC5078 & 170 & $<0.001$ & 55 & 0.002 & 28 & 0.004 & 88 & 0.002 & 6.52e+36 & 10.978 \\
ESO444-011 & 171 & $<0.001$ & 111 & $<0.001$ & 120 & $<0.001$ & 176 & $<0.001$ & 7.74e+35 & 8.824 \\
ESO508-039 & 172 & $<0.001$ & 151 & $<0.001$ & 158 & $<0.001$ & 179 & $<0.001$ & 1.09e+35 & 7.434 \\
NGC4748 & 173 & $<0.001$ & 82 & 0.001 & 37 & 0.002 & 115 & 0.001 & 2.32e+36 & 10.672 \\
J132044.59-302043.7 & 174 & $<0.001$ & -- & -- & -- & -- & 181 & $<0.001$ & -- & -- \\
ESO444-012 & 175 & $<0.001$ & 78 & 0.001 & 56 & 0.001 & 137 & $<0.001$ & 3.17e+36 & 10.26 \\
PGC141595 & 176 & $<0.001$ & 141 & $<0.001$ & 121 & $<0.001$ & 180 & $<0.001$ & 2.06e+35 & 8.881 \\
PGC850539 & 177 & $<0.001$ & 122 & $<0.001$ & 127 & $<0.001$ & 183 & $<0.001$ & 5.31e+35 & 8.668 \\
PGC740755 & 178 & $<0.001$ & 132 & $<0.001$ & -- & -- & 184 & $<0.001$ & 3.76e+35 & -- \\
ESO444-002 & 179 & $<0.001$ & 157 & $<0.001$ & 161 & $<0.001$ & 186 & $<0.001$ & 1.03e+35 & 7.375 \\
PGC763675 & 180 & $<0.001$ & 159 & $<0.001$ & 155 & $<0.001$ & 187 & $<0.001$ & 6.22e+34 & 7.643 \\
NGC5124 & 181 & $<0.001$ & 62 & 0.001 & 30 & 0.003 & 91 & 0.002 & 6.38e+36 & 11.025 \\
ESO444-033 & 182 & $<0.001$ & 140 & $<0.001$ & 147 & $<0.001$ & 188 & $<0.001$ & 2.69e+35 & 8.095 \\
WINGSJ132507.84-315046.2 & 183 & $<0.001$ & -- & -- & -- & -- & 189 & $<0.001$ & -- & -- \\
WINGSJ132507.85-315046.2 & 184 & $<0.001$ & 175 & $<0.001$ & -- & -- & 190 & $<0.001$ & 6.71e+33 & -- \\
PGC141602 & 185 & $<0.001$ & 138 & $<0.001$ & 143 & $<0.001$ & 191 & $<0.001$ & 3.89e+35 & 8.239 \\
NGC5048 & 186 & $<0.001$ & 74 & 0.001 & 44 & 0.001 & 128 & 0.001 & 5.21e+36 & 10.713 \\
PGC732248 & 187 & $<0.001$ & 143 & $<0.001$ & 139 & $<0.001$ & 192 & $<0.001$ & 2.49e+35 & 8.403 \\
SDSSJ120133.99+042759.3 & 188 & $<0.001$ & 160 & $<0.001$ & 137 & $<0.001$ & 194 & $<0.001$ & 7.38e+34 & 8.456 \\
NGC5051 & 189 & $<0.001$ & 76 & 0.001 & 39 & 0.002 & 119 & 0.001 & 4.78e+36 & 10.85 \\
PGC046903 & 190 & $<0.001$ & 150 & $<0.001$ & -- & -- & 195 & $<0.001$ & 1.95e+35 & -- \\
IC0879 & 191 & $<0.001$ & 147 & $<0.001$ & 115 & $<0.001$ & 193 & $<0.001$ & 2.54e+35 & 9.272 \\
PGC141596 & 192 & $<0.001$ & 121 & $<0.001$ & 85 & $<0.001$ & 182 & $<0.001$ & 9.42e+35 & 9.764 \\
PGC722221 & 193 & $<0.001$ & 170 & $<0.001$ & 168 & $<0.001$ & 198 & $<0.001$ & 2.70e+34 & 7.113 \\
PGC117211 & 194 & $<0.001$ & 162 & $<0.001$ & 136 & $<0.001$ & 197 & $<0.001$ & 8.52e+34 & 8.687 \\
ESO444-015 & 195 & $<0.001$ & 107 & $<0.001$ & 75 & $<0.001$ & 178 & $<0.001$ & 2.42e+36 & 10.053 \\
HIPASSJ1457-67 & 196 & $<0.001$ & -- & -- & 133 & $<0.001$ & 196 & $<0.001$ & -- & 8.823 \\
PGC042964 & 197 & $<0.001$ & 166 & $<0.001$ & 165 & $<0.001$ & 199 & $<0.001$ & 6.01e+34 & 7.317 \\
ESO575-041 & 198 & $<0.001$ & 127 & $<0.001$ & 135 & $<0.001$ & 200 & $<0.001$ & 1.20e+36 & 8.879 \\
PGC939548 & 199 & $<0.001$ & 149 & $<0.001$ & 141 & $<0.001$ & 201 & $<0.001$ & 4.06e+35 & 8.62 \\
PGC1295846 & 200 & $<0.001$ & 161 & $<0.001$ & 145 & $<0.001$ & 203 & $<0.001$ & 1.76e+35 & 8.642 \\
PGC2793691 & 201 & $<0.001$ & 142 & $<0.001$ & -- & -- & 204 & $<0.001$ & 7.10e+35 & -- \\
ESO443-086 & 202 & $<0.001$ & 124 & $<0.001$ & 124 & $<0.001$ & 202 & $<0.001$ & 1.88e+36 & 9.436 \\
PGC043964 & 203 & $<0.001$ & 120 & $<0.001$ & 77 & $<0.001$ & 185 & $<0.001$ & 2.56e+36 & 10.361 \\
NGC5126 & 204 & $<0.001$ & 109 & $<0.001$ & 60 & $<0.001$ & 157 & $<0.001$ & 4.67e+36 & 10.816 \\
PGC104887 & 205 & $<0.001$ & 169 & $<0.001$ & 149 & $<0.001$ & 205 & $<0.001$ & 8.69e+34 & 8.616 \\
\hline
% \end{tabular}
\end{longtable}

% Don't change these lines
%\bsp	% typesetting comment
\label{lastpage}

\end{document}